\documentclass[journal]{IEEEtran}
\usepackage{cite}
  \usepackage[pdftex]{graphicx}
\usepackage{amsmath}
\usepackage[caption=false,font=normalsize,labelfont=sf,textfont=sf, position=top]{subfig}
\usepackage[utf8]{inputenc}
\usepackage{epsf, amssymb}
\usepackage{amsthm}
\usepackage{xcolor, booktabs}
\usepackage{pdflscape}
\usepackage{hyperref, enumitem, rotating}
\usepackage{mathtools}
\usepackage{multicol, multirow}
\usepackage{algpseudocode}
\usepackage{accents}
\usepackage[export]{adjustbox}

\begin{document}
%
% paper title
% Titles are generally capitalized except for words such as a, an, and, as,
% at, but, by, for, in, nor, of, on, or, the, to and up, which are usually
% not capitalized unless they are the first or last word of the title.
% Linebreaks \\ can be used within to get better formatting as desired.
% Do not put math or special symbols in the title.
\title{A dynamic extreme value model \\ with applications to volcanic eruption forecasting}
%
%
% author names and IEEE memberships
% note positions of commas and nonbreaking spaces ( ~ ) LaTeX will not break
% a structure at a ~ so this keeps an author's name from being broken across
% two lines.
% use \thanks{} to gain access to the first footnote area
% a separate \thanks must be used for each paragraph as LaTeX2e's \thanks
% was not built to handle multiple paragraphs
% Use e.g. \IEEEmembership{Life~Fellow,~IEEE} if needed.

\author{Michele~Nguyen,~Almut~E.~D.~Veraart,~Benoit~Taisne,~Tan~Chiou~Ting,~and~David~Lallemant~% <-this % stops a space
\thanks{M. Nguyen is with the Asian School of the Environment, Nanyang Technological University (NTU), Singapore 637459 (email: michele.nguyen@ntu.edu.sg).}% <-this % stops a space
\thanks{A. E. D. Veraart is with the Department of Mathematics, Imperial College London, London SW7 2AZ, United Kingdom.}
\thanks{B. Taisne and D. Lallemant are with the Asian School of Environment and also with the Earth Observatory of Singapore, NTU.}% <-this % stops a space
\thanks{Tan Chiou Ting was with the Earth Observatory of Singapore, NTU.}
%\thanks{Manuscript received April 19, 2005; revised August 26, 2015.}
}

% note the % following the last \IEEEmembership and also \thanks - 
% these prevent an unwanted space from occurring between the last author name
% and the end of the author line. i.e., if you had this:
% 
% \author{....lastname \thanks{...} \thanks{...} }
%                     ^------------^------------^----Do not want these spaces!
%
% a space would be appended to the last name and could cause every name on that
% line to be shifted left slightly. This is one of those "LaTeX things". For
% instance, "\textbf{A} \textbf{B}" will typeset as "A B" not "AB". To get
% "AB" then you have to do: "\textbf{A}\textbf{B}"
% \thanks is no different in this regard, so shield the last } of each \thanks
% that ends a line with a % and do not let a space in before the next \thanks.
% Spaces after \IEEEmembership other than the last one are OK (and needed) as
% you are supposed to have spaces between the names. For what it is worth,
% this is a minor point as most people would not even notice if the said evil
% space somehow managed to creep in.

% The paper headers
% E.g. Journal of \LaTeX\ Class Files,~Vol.~14, No.~8, August~2015
\markboth{}%
{Nguyen \MakeLowercase{\textit{et al.}}: A dynamic extreme value model with applications to volcanic eruption forecasting}
% The only time the second header will appear is for the odd numbered pages
% after the title page when using the twoside option.
% 
% *** Note that you probably will NOT want to include the author's ***
% *** name in the headers of peer review papers.                   ***
% You can use \ifCLASSOPTIONpeerreview for conditional compilation here if
% you desire.

% If you want to put a publisher's ID mark on the page you can do it like
% this:
%\IEEEpubid{0000--0000/00\$00.00~\copyright~2015 IEEE}
% Remember, if you use this you must call \IEEEpubidadjcol in the second
% column for its text to clear the IEEEpubid mark.

% use for special paper notices
%\IEEEspecialpapernotice{(Invited Paper)}

% make the title area
\maketitle

% As a general rule, do not put math, special symbols or citations
% in the abstract or keywords.
\begin{abstract}
Extreme events such as natural and economic disasters leave lasting impacts on society and motivate the analysis of extremes from data. While classical statistical tools based on Gaussian distributions focus on average behaviour and can lead to persistent biases when estimating extremes, extreme value theory (EVT) provides the mathematical foundations to accurately characterise extremes. In this paper, we adapt a dynamic extreme value model recently introduced to forecast financial risk from high frequency data to the context of natural hazard forecasting. We demonstrate its wide applicability and flexibility using a case study of the Piton de la Fournaise volcano. The value of using EVT-informed thresholds to identify and model extreme events is shown through forecast performance. 
\end{abstract}

% Note that keywords are not normally used for peerreview papers.

% For peer review papers, you can put extra information on the cover
% page as needed:
% \ifCLASSOPTIONpeerreview
% \begin{center} \bfseries EDICS Category: 3-BBND \end{center}
% \fi
%
% For peerreview papers, this IEEEtran command inserts a page break and
% creates the second title. It will be ignored for other modes.
\IEEEpeerreviewmaketitle

\section{Introduction} \label{sec:intro}
% The very first letter is a 2 line initial drop letter followed
% by the rest of the first word in caps.
% 
% form to use if the first word consists of a single letter:
% \IEEEPARstart{A}{demo} file is ....
% 
% form to use if you need the single drop letter followed by
% normal text (unknown if ever used by the IEEE):
% \IEEEPARstart{A}{}demo file is ....
% 
% Some journals put the first two words in caps:
% \IEEEPARstart{T}{his demo} file is ....
% 
% Here we have the typical use of a "T" for an initial drop letter
% and "HIS" in caps to complete the first word.
\IEEEPARstart{N}{atural} hazards and extreme financial loss can be seen as extreme events, i.e. events which have low probabilities of occurring under normal circumstances. Due to their imbalanced number of occurrences, we usually have more data on common, non-extreme events than on extreme events. However, it can be shown that fitting a distribution to all the data via classical statistical methods leads to reasonable fit to the bulk of the data at the expense of poor fit to the tails in which the extremes lie \cite{ribatet2016}. Extreme value theory (EVT) and its corresponding models seek to remedy this disparity by offering guidance on when an extreme regime kicks in and how it can be modelled (see \cite{embrechts2013, beirlant2004}). For example, under known conditions, the distribution of excesses above a fixed threshold can be shown to converge to a Generalised Pareto Distribution (GPD) as the threshold increases. By knowing this asymptotic distribution, we can conduct model checks and generate sensible estimates of extremal behaviour.
\\
EVT is increasingly used in financial applications and was shown to give more accurate tail-risk predictions \cite{Danielsson1997, Longin2000}. To address the fact that there is time dependence in financial returns which goes against the traditional assumption of independence \cite{diebold1998}, \cite{Bee2019} propose a dynamic extreme value model which uses high-frequency realized measures of the daily asset price variation as covariates to model the probability of exceeding a high threshold and the size of the excesses. Since the realized variation are time-varying, the estimates of threshold exceedance and excesses are also time-varying.
\\
The extreme value model developed by \cite{Bee2019} can be adapted to wider settings. In this paper, we demonstrate this by adapting the model for natural hazard forecasting, specifically for volcanic eruptions. Just as how we can define extreme loss as an exceedance over some financial threshold, we can define extreme volcanic activity as the threshold exceedance of some, e.g. energy, index. The contributions of this model to the eruption forecasting literature are manifold. While a short overview of existing methods is provided in the Supplementary Information, we highlight the key differences and contributions in the next section.

\subsection{Contributions of the proposed model}

 The full and varied potential of machine learning algorithms for short-term eruption forecasting has yet to be realised \cite{Malfante2018, carniel2020, whitehead2021}. The proposed dynamic extreme value model draws on techniques from machine learning, time series analysis and EVT. To adapt it from its original financial context to wider settings, we need to decide on:
\begin{enumerate}
    \item a suitable index to compute threshold exceedances;
    \item the look-ahead window or forecast horizon;
    \item the auxilliary information or covariates used to inform future behaviour;
    \item and the time periods which we compute these from, i.e. the covariate window. 
\end{enumerate}

In the seismic context, threshold exceedance for event detection is synonymous with first arrival picking algorithms such the classical short time average over long time average (STA/LTA) method and a recently introduced method using trace envelopes \cite{Withers1998, Trnkoczy1999, Al-Mashhor2019}. Since trace envelopes can be interpreted as the amount of energy in the signal, in our demonstration, we will use them as eruption indices to take threshold exceedances of. These exceedances would hopefully relate to extreme regimes leading up to volcanic eruptions and can be forecasted using covariates. Note that by modelling the exceedance of an eruption index rather than raw monitoring signal (as done by \cite{sobradelo2015} via a Bayesian event tree), we avoid the need to interpret estimates in real-time. Since different frequency bands within the seismic signal represent different physical phenomena \cite{bormann2013}, we will consider envelopes of frequency-filtered data. 
\\
While some existing techniques like the failure forecast method estimate the eruption onset time directly, others including event trees define a look-forward window within which we estimate the probability of an eruption occurring. We take the latter approach. In particular, for illustration, we produce one-hour ahead forecasts to complement other longer term forecasts. Although a longer lead-time allows for more time to make emergency management decisions, notify the public and implement evacuations \cite{wild2021}, forecasts are typically more accurate when made closer to the actual eruption time due to temporal divergence at larger lags (see for example, \cite{Sugihara1990}). 
\\
Eruption forecasting methods such as event trees, belief networks and process/source models presuppose precursors or associations between source mechanisms and time series signals  \cite{Brenguier2008}. In contrast, our proposed methodology selects combinations of covariates which could represent precursors for any volcano and type of eruption if we train the model on corresponding data. Specifically, we test covariates inspired by machine learning classification algorithms for seismic signals \cite{Malfante2018}. By combining these covariates, different aspects of the seismic data and relationships across different frequency bands are represented. An objective stepwise selection procedure is then used to determine which covariates are more informative for forecasting eruptions.

\subsection{Paper structure}

We continue in \ref{sec:model} by outlining the key components of the dynamic extreme value model introduced by \cite{Bee2019}. In Section \ref{sec:casestudy}, we introduce our case study, the Piton de la Fournaise volcano, and illustrate how the model can be adapted for eruption forecasting. By comparing the effect of different choices of the threshold on the model fit and training performance, we highlight the value of using EVT to guide threshold choice in Section \ref{sec:EVTvalue}. In Section \ref{sec:forecasteval}, we evaluate the broader applicability of the method by refitting the model using three training event sets and testing the calibrated model on both event and non-event sets. In Section \ref{sec:discussion}, we conclude by discussing our results and suggesting future areas for research. The code used for the analysis is publicly available at \url{https://github.com/ntu-dasl-sg/dynamic-EV-forecasting}. 

\section{The dynamic extreme value model} \label{sec:model}

Following \cite{Bee2019}, let $\{Y_{t}\}_{t = 1, \dots, T}$ denote a time series of an index where higher values are associated with extreme events.  Based on the selected threshold $u \in \mathbb{R}$ of the index, we define \textit{exceedances} to be the binary indicators of whether the index is higher than the threshold and define the \textit{excesses} to be the numerical values of the exceedances, i.e. how much we exceed by. The conditional probability that the index at time $t$, $Y_{t}$, exceeds $u$ by some excess $z>0$ given prior information available at time $t$, $\mathcal{F}_{t-1}$, can be written as:
\begin{align}
    &P(Y_{t} > u + z | \mathcal{F}_{t-1}) \nonumber \\
    =& P(Y_{t} > u|\mathcal{F}_{t-1})P(Y_{t}-u>z | Y_{t}>u, \mathcal{F}_{t-1}) \nonumber \\
    =& \phi_{t}|\mathcal{F}_{t-1} \times GPD(\xi_{t}, \nu_{t})|\mathcal{F}_{t-1}. \label{eqn:condmod}
\end{align}

Here, $\phi_{t}|\mathcal{F}_{t-1} =  P(Y_{t} > u|\mathcal{F}_{t-1})$ represents a time-varying, Binomial exceedance probability which can be modelled with a logit function:
\begin{align}
    \phi_{t}|\mathcal{F}_{t-1} &= \frac{\exp(\psi_{0} + \sum_{i = 1}^{p}\psi_{i}x^{(i)}_{t-1})}{1 + \exp(\psi_{0} + \sum_{i = 1}^{p}\psi_{i}x^{(i)}_{t-1})}, \label{eqn:logitphi}
\end{align}

where $\mathbf{x}_{t-1} = (x_{t-1}^{(1)}, 
\dots, x_{t-1}^{(p)})$ denotes a vector of $p$ covariates from the previous time step and is used to project the future probability. The parameters $\{\psi_{i}\}_{i = 0, \dots, p}$ can be estimated by maximising the likelihood function:
%For financial asset prices in \cite{Bee2019}, the covariates used were realised measures of variation e.g. 5-min realised variation, realised jumps measure to disentangle the contribution of jumps to the realized variation from that of the Brownian path, realized measures sampled at a frequency lower than 5-min to assess the effect of microstructure noise and noise-robust realized measures.
\begin{align}
    \mathcal{L}(\boldsymbol{\psi}; I_{t}, \mathbf{x}_{t}) =& \prod_{t = l+1}^{T} \left(\exp(\psi_{0} + \sum_{i = 1}^{p}\psi_{i}x^{(i)}_{t-1})\right)^{I_{t}} \nonumber \\ 
    &\times \frac{1}{1 + \exp(\psi_{0} + \sum_{i = 1}^{p}\psi_{i}x^{(i)}_{t-1})},
    \label{eqn:te_regression}
\end{align}

where $l$ is the lag at which the covariates $\mathbf{x}_{t}$ become available and $I_{t}$ is the indicator of an exceedance at time $t$ (it is equal to $1$ if there is an exceedance and $0$ otherwise). This is equivalent to using a \textit{logistic regression} to model the probability of threshold exceedance.

In (\ref{eqn:condmod}), the model for excesses of the threshold is given by a Generalised Pareto distribution (GPD):
\begin{equation}
P(Y_{t}-u>z | Y_{t}>u, \mathcal{F}_{t-1}) =  GPD(\xi_{t}, \nu_{t})|\mathcal{F}_{t-1}.
\end{equation}

The shape parameter $\xi_{k} = \xi$ is estimated assuming a non-time varying GPD distribution for the excesses and kept constant for stability. To account for the time-varying nature of the excess distribution, we model the scale parameter $\nu_{t}$ as a log-linear function:
\begin{equation}
    \nu_{t} = \exp\left(\kappa_{0} + \sum_{i = 1}^{q}\kappa_{i}x^{(i)}_{t-1}\right).
\end{equation}

When $\kappa_{i} = 0$ for $i>0$, this means that it is sufficient to model the distribution of the excesses statically. From the definition of a GPD with non-zero shape parameter:
\begin{align}
&P(Y_{t}-u>z | Y_{t}>u, \mathcal{F}_{t-1}) \nonumber \\
=&  \left(1 + \frac{\xi z}{\exp(\kappa_{0} + \sum_{i = 1}^{q}\kappa_{i}x^{(i)}_{t-1})} \right)^{-1/\xi},
\end{align}

where $z = y_{t} - u$, the excess. If $\xi > 0$, $z\geq 0$ and if $\xi<0$, $0\leq z \leq -\exp(\kappa_{0} + \sum_{i = 1}^{q}\kappa_{i}x^{(i)}_{t-1})/\xi$, i.e. the excesses have an upper bound. The parameters $\{\kappa_{i}\}_{i = 1, \dots, q}$ can be estimated by maximising:
\begin{align}
    &\mathcal{L}(\mathbf{\kappa}, \xi; \mathbf{z}_{t}, \mathbf{x}_{t}) \nonumber \\
    =& \prod_{t = l + 1}^{T} \left(\frac{1}{\exp(\kappa_{0} + \sum_{i = 1}^{q}\kappa_{i}x^{(i)}_{t-1})} \right. \nonumber \\
    &\times \left.\left[\left(1 + \frac{\xi z_{t}}{\exp(\kappa_{0} + \sum_{i = 1}^{q}\kappa_{i}x^{(i)}_{t-1})} \right)^{-1/\xi-1}\right]_{+}\right)^{I_{t}}, \label{eqn:GPlik}
\end{align}

where $[x]_{+} = \max(0, x)$ and we add the time subscript to the excesses, $\mathbf{z}$, to denote their temporal indices. Henceforth, we shall refer to the maximisation of (\ref{eqn:GPlik}) to estimate $\{\kappa_{i}\}_{i = 1, \dots, q}$ as \textit{GPD regression}. When the estimated shape parameter $\hat{\xi}$ is not significantly different from zero, we set it equal to zero and use an \textit{exponential regression} instead.

%In \cite{Bee2019}, they prove the consistency and asymptotic normality of the maximum likelihood estimators under the assumption that the excesses $(z_{1}, \dots, z_{k})$ are independent with distribution (\ref{eqn:GPlik}) and $\xi \geq 0$. 

\begin{table*}[!t]
\centering
\begin{tabular}{l|c|ccc|cc}
\multicolumn{1}{l}{\multirow{2}{*}{}} & \multicolumn{1}{l}{}            & \multicolumn{3}{c}{Eruption}                                                                      & \multicolumn{2}{c}{Duration}                                            \\ 
\multicolumn{1}{l|}{}                  & \multicolumn{1}{c|}{Data period} & \multicolumn{1}{c}{Date (mm/dd/yyyy)} & \multicolumn{1}{c}{Location} & \multicolumn{1}{c|}{Type} & \multicolumn{1}{c}{Seismic crisis} & \multicolumn{1}{c}{Seismic swarm} \\ \hline 
\multicolumn{1}{l|}{}                  & \multicolumn{1}{l|}{} & \multicolumn{1}{l}{} & \multicolumn{1}{l}{} & \multicolumn{1}{l|}{} & \multicolumn{1}{l}{} & \multicolumn{1}{l}{} 
\\
Training event 1  & 11/04/2009 - 11/06/2009 & 11/05/2009        & Flank S SE & E-L  & 1h 30min       & 30min         \\
Training event 2  & 12/13/2009 - 12/14/2009 & 12/14/2009        & Flank S/SW & E-L  & 1h 10min       & 32min         \\
Training event 3  & 01/01/2010 - 01/03/2010 & 01/02/2010        & Dolomieu W & E-C  & 2h 30min       & 42min         \\
Test event        & 10/13/2010 - 10/14/2010 & 10/14/2010        & Flank S    & E-L  & 5h 30min       & 40min         \\
Test non-event 1  & 11/30/2009 & -                 & -          & -    & -              & -            \\
Test non-event 2  & 12/22/2009 - 12/23/2009 & -                 & -          & -    & -              & -            \\
Test non-event 3  & 05/08/2010 - 05/10/2010 & -                 & -          & -    & -              & -            \\
\hline \\
\end{tabular}
\caption{Events and non-events used to train and test the forecasting model. E-C refers to an eruption that remains inside the summit craters (Dolomieu or Bory) and E-L refers to an eruption that starts anywhere else outside of the summit craters.}
\label{table:events}
\end{table*}

\section{Case study: Piton de la Fournaise volcano} \label{sec:casestudy}

\subsection{Data}

We will use the dynamic extreme value model to forecast eruptions at the Piton de la Fournaise volcano. Situated on La Réunion Island, Piton de la Fournaise is one of the most active basaltic volcanoes, with an average of one eruption every 10 months \cite{Roult2012}. In addition to the existing seismic monitoring stations, 15 broad-band stations have been installed on the volcano as part of the Understanding Volcano project in 2009–2010 \cite{Taisne2011}. The data collected is available at \url{https://www.fdsn.org/networks/detail/YA_2009/}. For our analysis, we use data for four eruption events and three non-events obtained at the UV05 station between 2009 and 2010. This was the closest available station to the January 2010 eruptive center \cite{Journeau2020}. 
\\
Table \ref{table:events} shows the key characteristics of these events as outlined in \cite{Roult2012}. To evaluate the forecast performance in Section \ref{sec:forecasteval}, we will use three of the events for training the model and test it on another event as well as three non-events. Although an eruption also occurred on December 9th 2010, we did not use this data for training or testing because the location of the eruption (Flank N) is relatively far away from the rest of event locations and hence could skew our results. The non-event dates were chosen to be roughly halfway between the selected four events and represent quiet periods for which we should not expect threshold exceedances of our eruption indices.

%Figure \ref{fig:uv05} shows the position of seismic stations on Piton de la Fournaise. 
%An eruption at Piton de la Fournaise is usually preceded by an unrest period that may last days or months and is characterized by a slowly increasing seismicity and ground deformation \cite{Roult2012}. Immediate pre-eruptive crises are usually less than two hours, which is shorter than that of other volcanoes \cite{aki2000}. 
%Our data comes from the UV05 station which was the closest available station to the January 2010 eruptive center \cite{Journeau2020}. 
%\begin{figure}[!t]
%\includegraphics[width=3.5in]{figures/map_UV05.png}
%\caption{Location of the UV05 station on Piton de La Fournaise %from \cite{Taisne2011}}
%\label{fig:uv05}
%\end{figure}

\subsection{Illustration of method}

\begin{figure*}[!t]
\centering
\subfloat[Signal]{\includegraphics[width=3.5in,valign=t, trim = {0.8cm 0 0cm 0}]{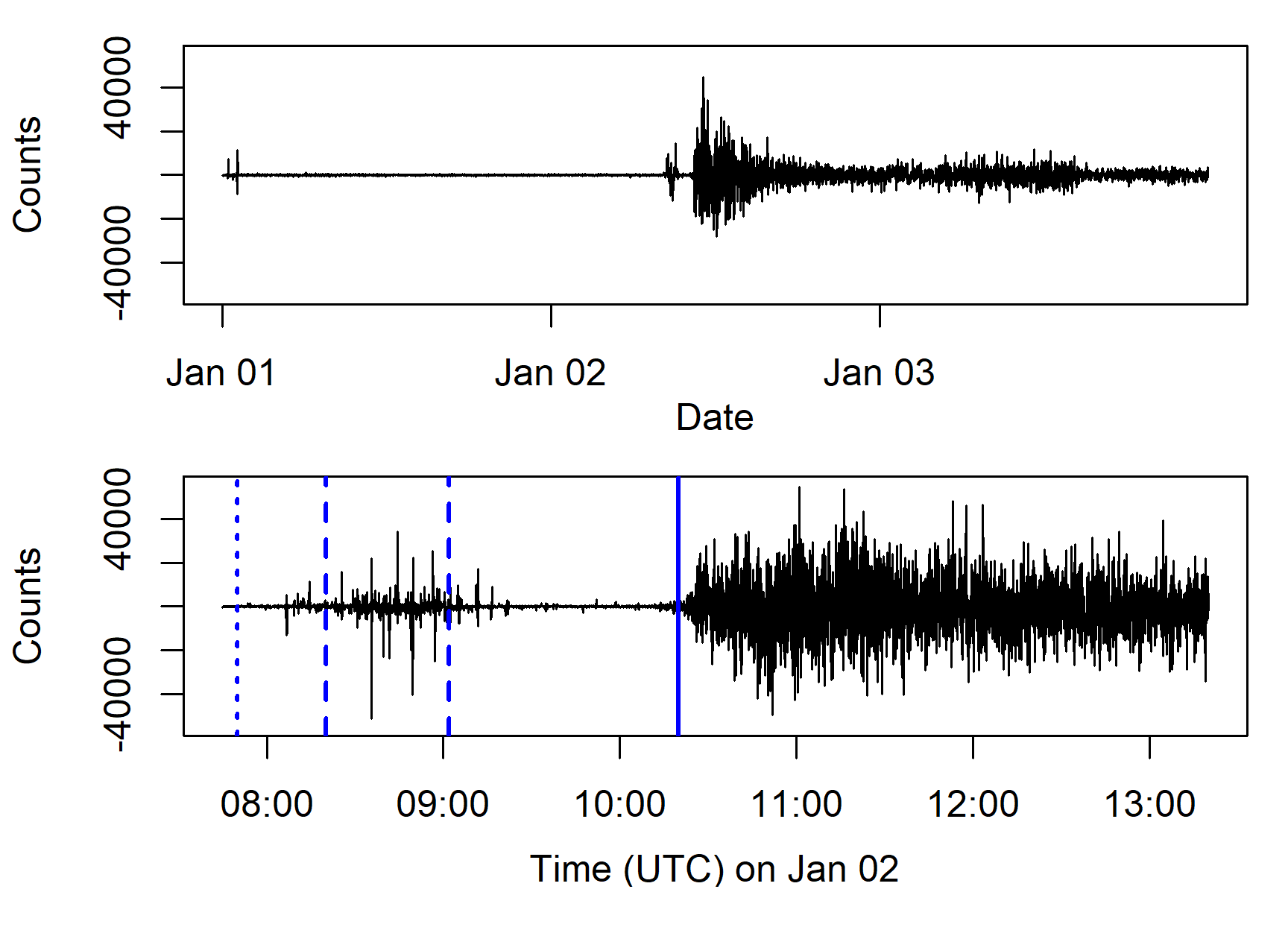}%
\label{fig:piton15signal}}
\hfil
\subfloat[Envelope index]{\includegraphics[width=3.5in,valign=t, trim = {0cm 0 0.8cm 0}]{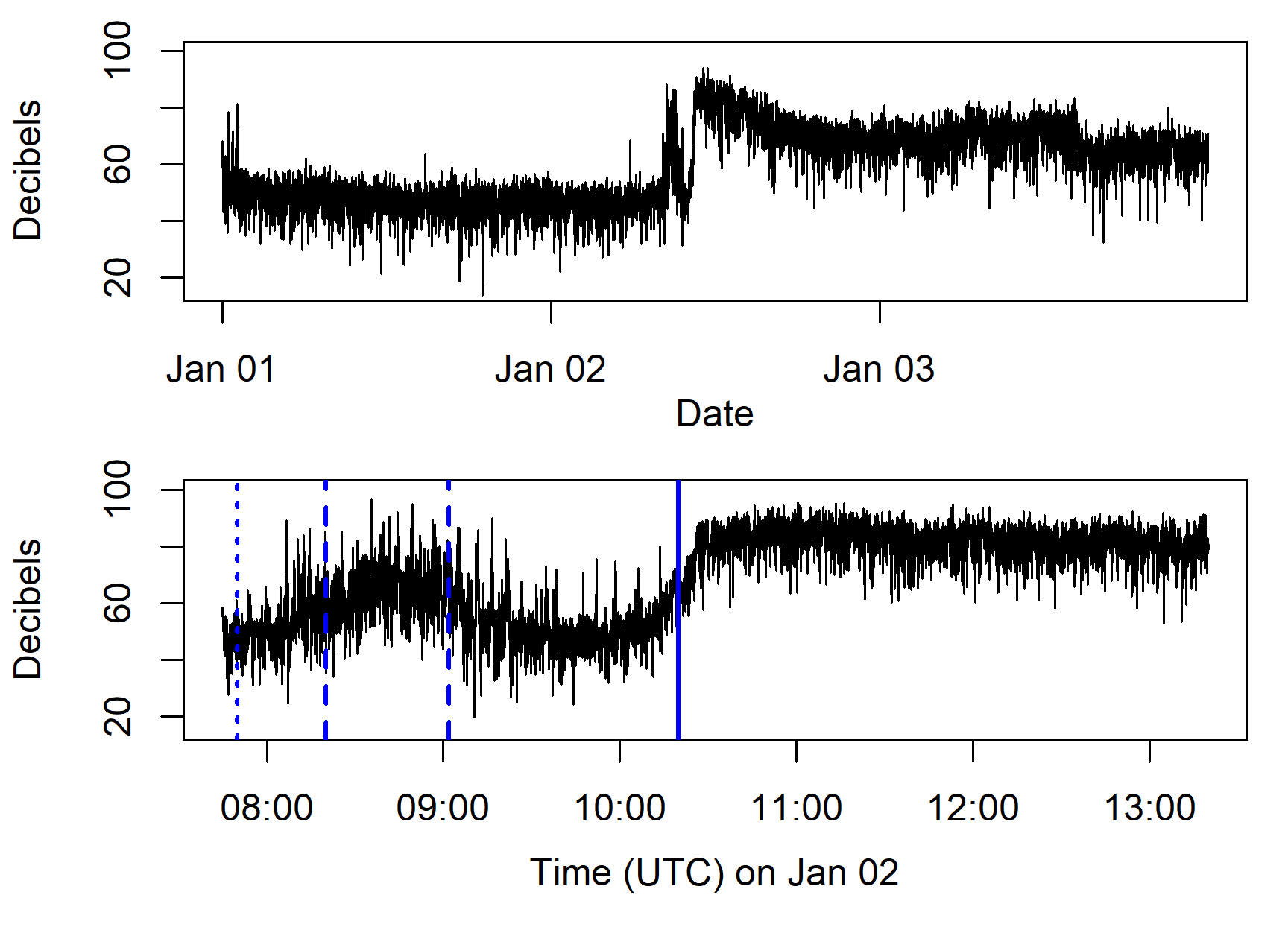}%
\label{fig_sampleindex}}
\\
\subfloat[Forecasting exceedances]{\includegraphics[width=3.8in,valign=t, trim = {2.5cm 0 4cm 0}]{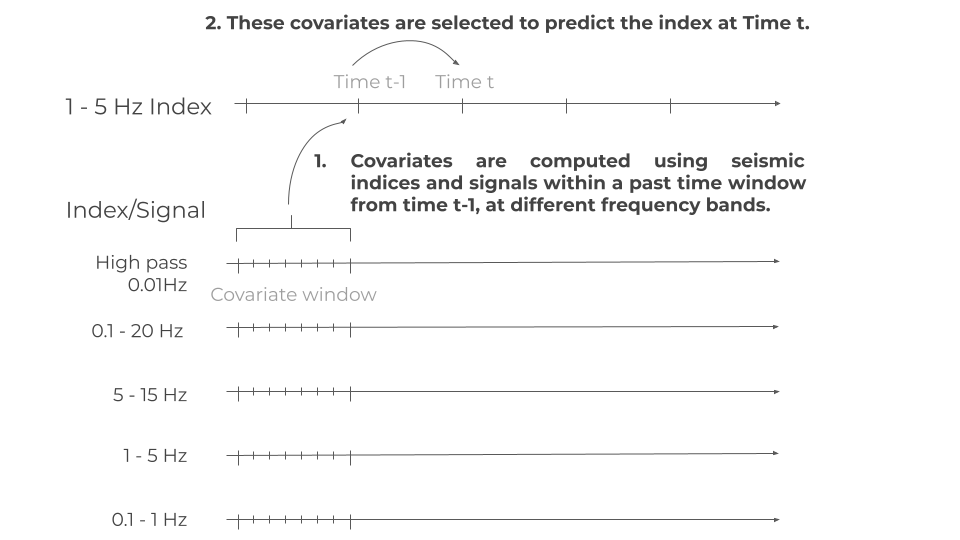}
\label{fig_forecasting}}
\subfloat[Choosing a threshold]{\includegraphics[width=3.2in,valign=t]{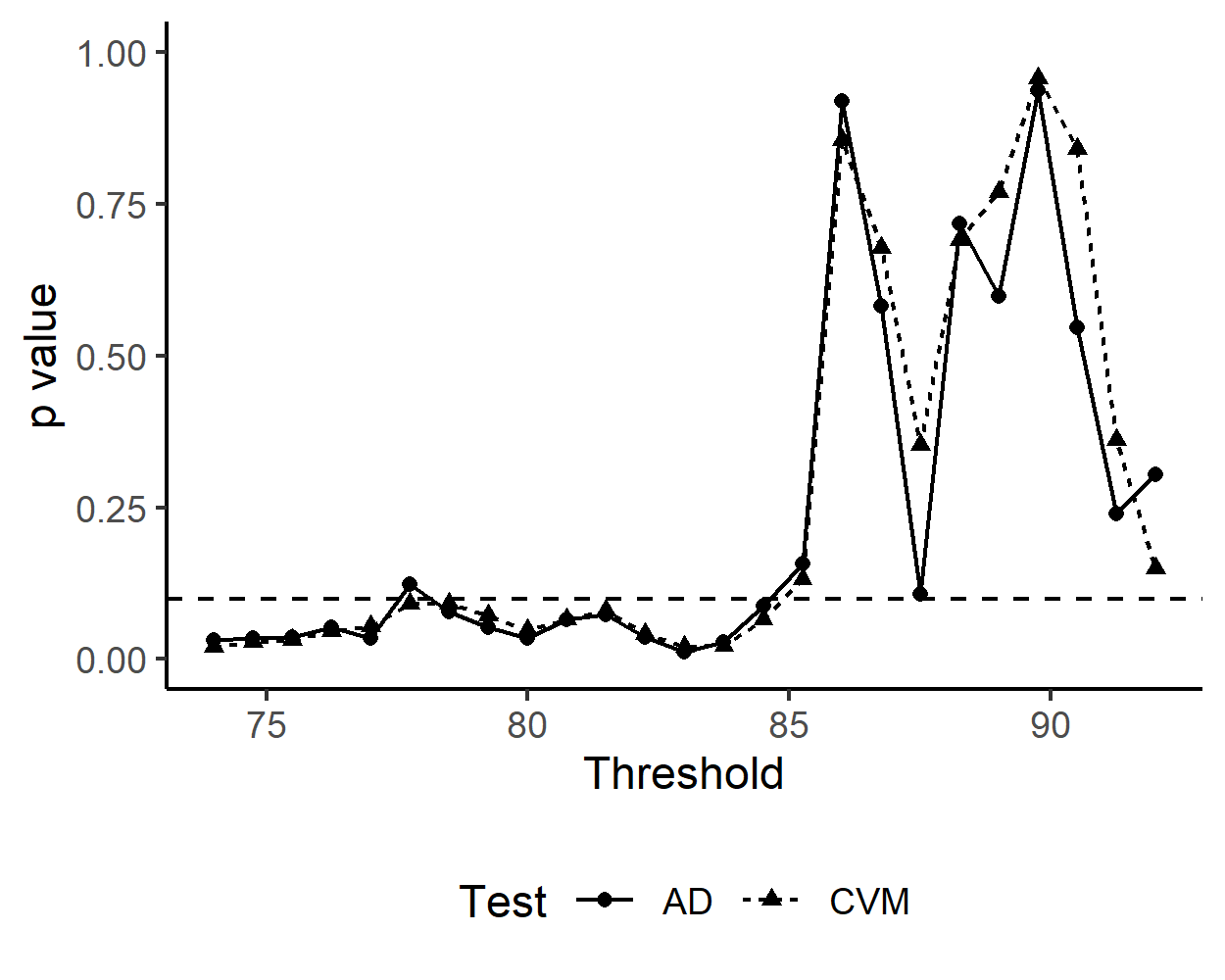}
\label{fig_gpd_pval}}

\caption{(a) Seismic signal for the 1-5Hz frequency band and (b) the corresponding envelope index in decibels (dB) for the January 2010 eruption event. The dotted, dashed and bold blue vertical lines denote the start of the seismic crisis, the start and end of the seismic swarm and the eruption onset respectively. Note that the time series are deciminated such as every 5183th reading and every 402th reading are shown for the top and bottom plots respectively. (c) Model conceptual diagram for the forecasting of the exceedance of the 1-5Hz envelope index based on covariates computed from signals and indices across multiple frequency bands within past time windows. (d) p-values from the Anderson-Darling (AD) and Cram{\'e}r-von Mises (CVM) tests for goodness-of-fit of the excesses to the GPD distribution (we select the lowest threshold for which the p-value exceeds the 10\% significance level). }
\label{fig_methodillustration}
\end{figure*}

To illustrate the use of the dynamic extreme value model for eruption forecasting, we will first apply the method to the January 2010 event data (Training event 3) since it is best documented out of the chosen events. Figure \ref{fig:piton15signal} shows the raw seismic signal for the 1-5Hz frequency-filtered data. The bottom plot focuses on January 2nd when the eruption occurred and the dotted, dashed and bold blue vertical lines denote the start and end timings of the seismic crisis, swarm and eruption onset respectively. As documented in \cite{Roult2012}, a seismic crisis took place from 07:50 local time. Between 08:10 and 09:02, a seismic swarm with high level of seismicity was recorded. This is reflected in the seismic signal as larger fluctuations in the readings between the dashed blue vertical lines. The swarm was followed by a relatively quiet phase that directly preceded the onset of the eruption at about 10:20 (the eruption is indicated by the continuous seismic tremor in the figure). The whole eruption was estimated to have lasted 9.6 days, ending at 00:05 on January 12th 2010.

\subsubsection{Trace envelope as an eruption index}Before fitting the dynamic extreme value model for eruption forecasting, we need to decide on eruption index which we want to consider threshold exceedances of. We propose to use the trace envelope $E_{t}$ for $t = 1, \dots, T$. This can be seen as the instantaneous amplitude of the seismic trace $\mathbf{s} = (s_{1}, \dots, s_{T})$, and can be computed as follows:

\begin{enumerate}
    \item First, we compute the discrete Fourier transform (DFT) of $\mathbf{s}$ (this is implemented by the R function `fft') for $t \in \{1, \dots, T\}$:
   \begin{equation}
       f_{t} = DFT(\mathbf{s})_{T} = \sum_{k=1}^{T} s_{k}\exp(-2\pi i(k-1)(t-1)/T),
   \end{equation} 
   where $T$ is the length of the seismic trace. Set $\mathbf{f} = (f_{1}, \dots, f_{T})$. 
   \item Next, we compute the complex Hilbert fast Fourier transform (FFT) of $\mathbf{s}$ as:
   \begin{equation}
       H_{t} = IFT(\mathbf{f}\mathbf{h})/T, 
   \end{equation}
   where the length-$T$ series $\mathbf{h} = (1, 2, 2, \dots, 2, 2, 1)$ if $T$ is even and $(1, 2, 2, \dots, 2, 2, 2)$ if $T$ is odd and the inverse Fourier Transform (IFT) is defined as:
   \begin{equation}
      IFT(\mathbf{f}\mathbf{h})_{t} = \sum_{k=1}^{T} f_{k}h_{k}\exp(2\pi i(k-1)(t-1)/T),
   \end{equation} 
   for $t \in \{1, \dots, T\}$.  
 % Note that the real part of $H_{t}$, $Re(H_{t}) = s_{t}$, the seismic trace and the imaginary part $Im(H_{t})$ is its Hilbert transform. 
  \item For $t \in \{1, \dots, T\}$, the trace envelope of the seismic trace $\mathbf{s}$ is defined as:
   \begin{equation}
        E_{t} = \sqrt{Re^{2}(H_{t}) + Im^{2}(H_{t})} = Mod(H_{t}).
    \end{equation}
%Note that we have not removed any linear or non-zero mean of seismic trace before computing the envelope and have not used cosine tapers to avoid boundary effects. 
\end{enumerate}

The above steps are in line with those used within the R function `envelope' in the IRISSeismic package.
\\
In \cite{Al-Mashhor2019}, the first arrival travel time picking algorithm was based on the envelope in decibels. Thus, we use $Y_{t} = 20\log_{10}(E_{t})$ as our eruption index. Figure \ref{fig_sampleindex} shows the trace envelope time series computed from the 1-5Hz frequency-filtered data for the January 2010 event. We focus on the 1-5Hz frequency band because it is strongly associated with volcanic-tectonic activity. We see that although there are some spikes at the start of January 1st, the envelope index remains relatively low around 50 decibels until the recorded seismic crisis on January 2nd (represented by the dotted blue vertical line in the bottom plot). From the time of the seismic crisis, the index increases to a first peak during the seismic swarm before waning slightly. About 10:00, the index increases steadily to plateau slightly above 80 decibels, the timing of which coincides with the recorded eruption onset (represented by the bold blue vertical line in the bottom plot). The relationship between the increases in the index and the seismic events enable us to use threshold exceedances to forecast eruptions.

\subsubsection{Forecast horizon and covariate window}
In addition to computing the index to take exceedances of, we also need to decide on: 
\begin{itemize}
    \item The forecast horizon $\delta_{t}$: the time period between the time window where the covariates are computed and the forecast time.
    \item The time window within which past observations contribute to the covariates, $W$.
\end{itemize}

For illustration, we use $\delta_{t} = W = 1$ hour so that we make 1-hour ahead forecasts with one hour of past data to inform the covariates in the model. This mimics the settings in \cite{Malfante2018}, \cite{ren2020} and \cite{Brenguier2008} where either one hour long signals were classified or the covariates were generated by scanning a moving window of length one hour across the seismic signals. This framework is pictured in Figure \ref{fig_forecasting} where the time period between $t-1$ and $t$ is  $\delta_{t} = 1$ hour and the covariate window $W = 1$ hour.
\\
Although we use the original, high-frequency (100Hz) data at different frequency bands (0.1-1 Hz, 1-5 Hz, 5-15 Hz, 0.1-20Hz and high pass 0.01Hz) to compute the covariates, we chose to produce forecasts only every 10 seconds to reduce unnecessary computational burden. For a practically useful workflow, this should be longer than the time required to compute the covariates from the past hour and use the fitted model to generate the forecast. 

\subsubsection{Threshold selection}
Next, we select a suitable threshold to define the extreme regime to which we will associate the covariates and forecast extremal behaviour, i.e. exceedance. Based on the Anderson-Darling (AD) and Cram{\'e}r-von Mises (CVM) tests for goodness-of-fit of the excesses to the GPD distribution \cite{Barder2018}, Figure \ref{fig_gpd_pval} suggests that under a significance level of 10\%, a threshold of 85 decibels would be reasonable since this is the lowest value from the right for which the p-value exceeds the significance level. 
%This is further supported by the plateauing of the estimated GPD scale and shape parameters after the value of 85 in Figure \ref{fig:gpd_shape}. The method accounting for multiple testing in \cite{Barder2018} did not prove useful in this context because all ForwardStop p-values lay above the significance level for all considered thresholds. Figure \ref{fig:reenv_diagnostics} shows that the choice of $u = 85$ leads to a reasonable fit of the GPD. Note that this choice of a threshold gives us 423 exceedances in our dataset. Figure \ref{fig:threshold_on_index} shows that there are earlier exceedances prior and during the seismic swarm which are larger in magnitude than those during the eruption. This would have implications on the chosen covariates and predictions.

\subsubsection{Covariates}To forecast threshold exceedances of our eruption index, we compute the covariates suggested in \cite{Malfante2018} for our frequency-filtered data (0.1-1Hz, 1-5Hz, 5-15Hz, 0.1-20Hz and high pass 0.01Hz) and their associated trace envelopes. \cite{Malfante2018} introduce three domains of representation of the seismic time series which could be useful: the original temporal domain, the frequency domain where spectral content is obtained via a Fourier transform and the cepstral domain where we compute the Fourier transform twice to highlight the harmonic properties of a given signal. 
\\
For each of these three representations of the seismic traces and their trace envelopes, we can compute:
\begin{itemize}
    \item Statistical features: e.g. standard deviation, skewness, root mean square bandwidth, and kurtosis which captures the transition between two signals. 
    %and is calculated using a recursive formula which is stable in the presence of strong sets by \cite{Langet2014}), 
    \item Entropy features: e.g. Shannon entropy which describes the distribution of the amplitude levels of a given signal.
    \item Shape descriptor features: e.g. rate of attack (ROA) which is defined as $\max_{i}\left(\frac{s_{i}-s_{i-1}}{n}\right)$ and rate of decay which is defined as $\min_{i}\left(\frac{s_{i}-s_{i-1}}{n}\right)$ where $n$ is number of observations within the covariate window. %Note that this family of covariates is sensitive to outliers.
\end{itemize}

A feature or covariate can take on different meanings depending on the domain it is computed on. For example, the feature `i of Central Energy' which is the time around which the signal energy is centered or the time centroid in the temporal domain, can be interpreted as the fundamental frequency in the frequency domain and the harmonic frequency in the cepstral domain. In addition, while the ratio of the maximum value over mean value can describe the contrast and relate to the cause of the event in the original temporal domain, in the frequency domain, it describes the spectral richness of the signature and in the cepstral domain, harmonic content of an observation. 
%\\
%Compiling the above-mentioned covariates for the January 2010 event period took 3 hour and 20 minutes per frequency band on a PC with characteristics: Intel\textsuperscript{\textregistered} Xeon\textsuperscript{\textregistered} W-2112 CPU Processor @ 3.60GHz; 32GB of RAM; Windows 10 64-bit. 
%To implement this forecasting routine for longer time spans, one can also consider including covariates which are constructed based on longer-term and/or historical parameters instead of a short, 1 hour moving window. This could help borrow information from the past regarding longer term volcanic states. 
\\
To ensure that the model is not too sensitive towards extreme covariate values, the covariates were transformed before use. Box-Cox analyses were used to select which power or log-transformation was required to make their distributions more similar to Gaussian distributions. After tranformation, the covariates were standardised using their mean and standard deviations to be on similar scales. To account for multicollinearity, we ordered the covariates according to increasing Akaike Information Criterion (AIC) of their univariate models (for threshold exceedance and excesses separately). Then, we removed covariates which had more than 0.6 in absolute correlation to covariates which were deemed more informative than themselves.

\begin{figure*}[!t]
\centering
\subfloat[Training forecasts]{\includegraphics[width=3.5in,valign=t]{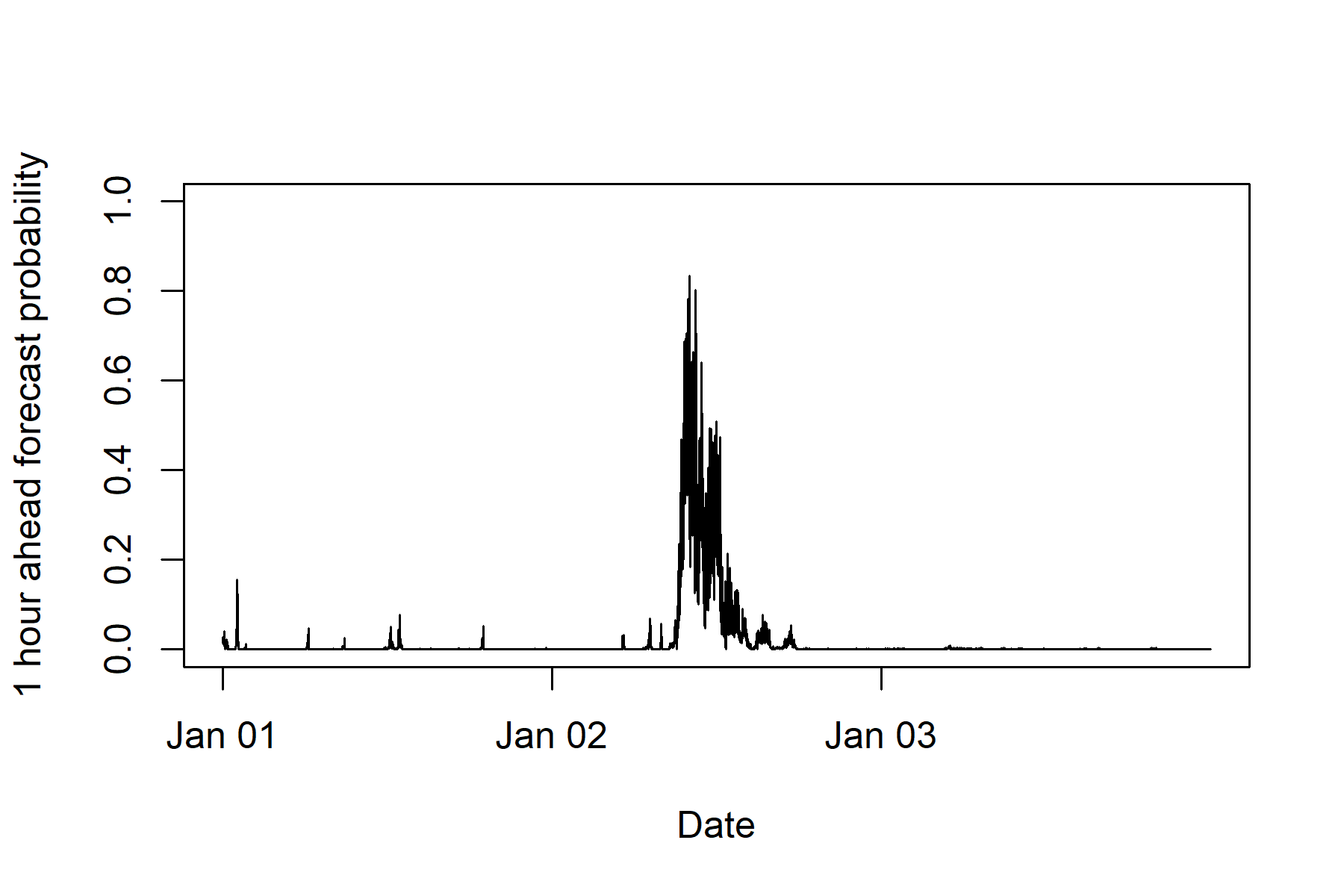}%
\label{fig:results_15_all}}
\hfil
\subfloat[Zoomed in]{\includegraphics[width=3.5in,valign=t]{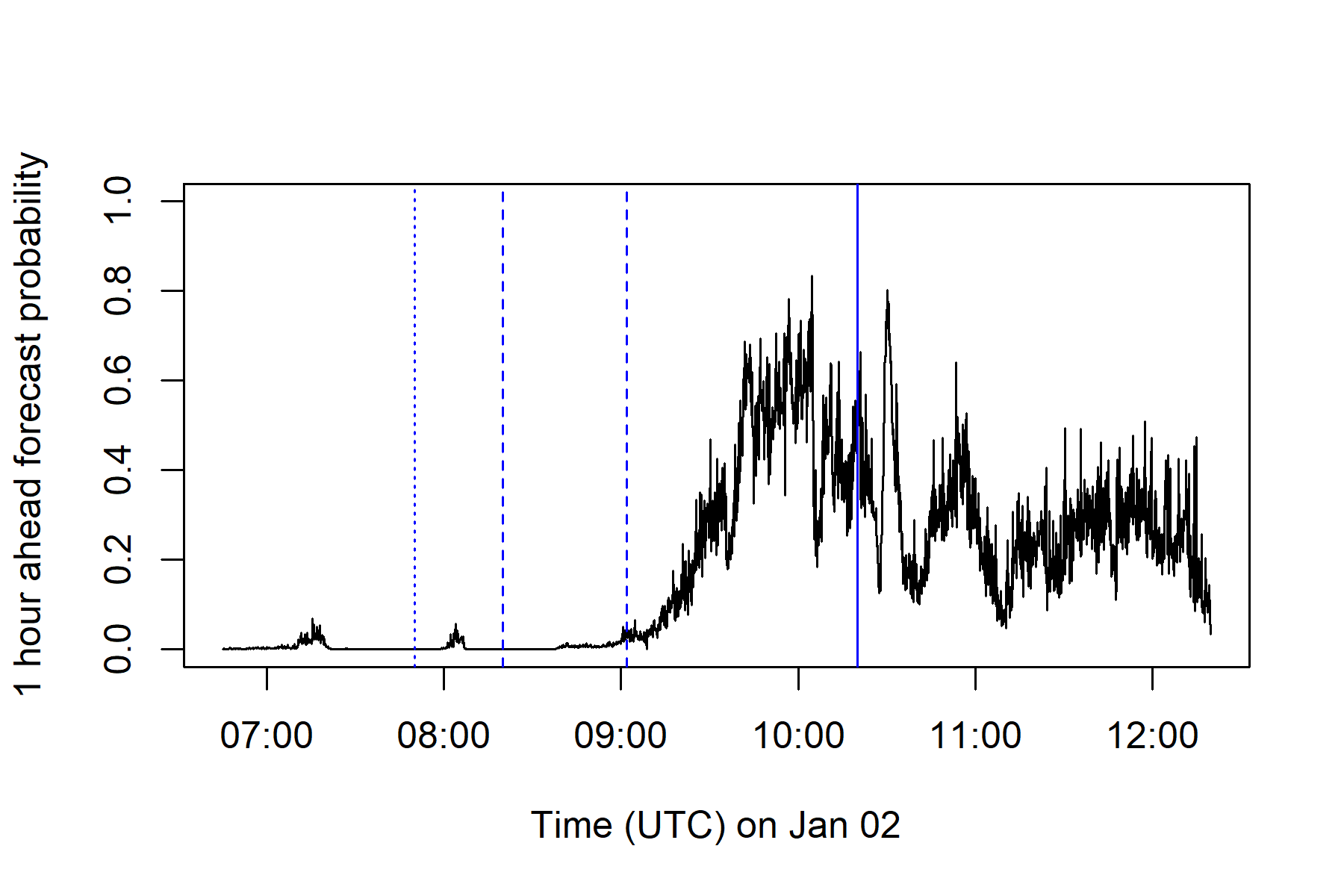}%
\label{fig:results_15_zoom}}

\caption{(a) Training one-hour-ahead exceedance forecasts based on the logistic regression for the 1-5Hz envelope index and the January 2010 eruption; (b) Zoomed into the period of significant volcanic activity (the dotted, dashed and bold blue vertical lines denote the start and end of the seismic crisis, swarm and eruption onset respectively). Here, the value of the black line at 09:00, for example, indicates the forecasted probability of exceedance for 10:00.}
\label{fig_results_15}
\end{figure*}

\subsubsection{Regression models} As outlined in Section \ref{sec:model}, we fit a logistic regression for threshold exceedances and an GPD regression for threshold excesses. The shape parameter of the GPD is fixed to the value of the estimate obtained using maximum likelihood estimation for a constant GPD. In this case, this has an asymptotically normal 95\% confidence interval of $(-0.300, -0.160)$ which does not include 0. The negative shape parameter implies that the distribution of the excesses is Pareto type II and lies within the Weibull domain of attraction which contains distributions with short tails, i.e. finite endpoints. 
\\
The models were chosen by stepwise selection based on AIC. For the logistic regression, the default choice of backwards followed by forward selection was used; for the GPD regression, we used forward before backwards selection to avoid singularity issues which arise from having large numbers of covariates and excess uninformative covariates in the model.
\\
Steps 1-5 can be repeated to model threshold exceedances and excesses for the other frequency-filtered envelopes.

\begin{figure*}[!t]
\centering
\subfloat[Goodness-of-fit]{\includegraphics[width=3.7in,valign=t]{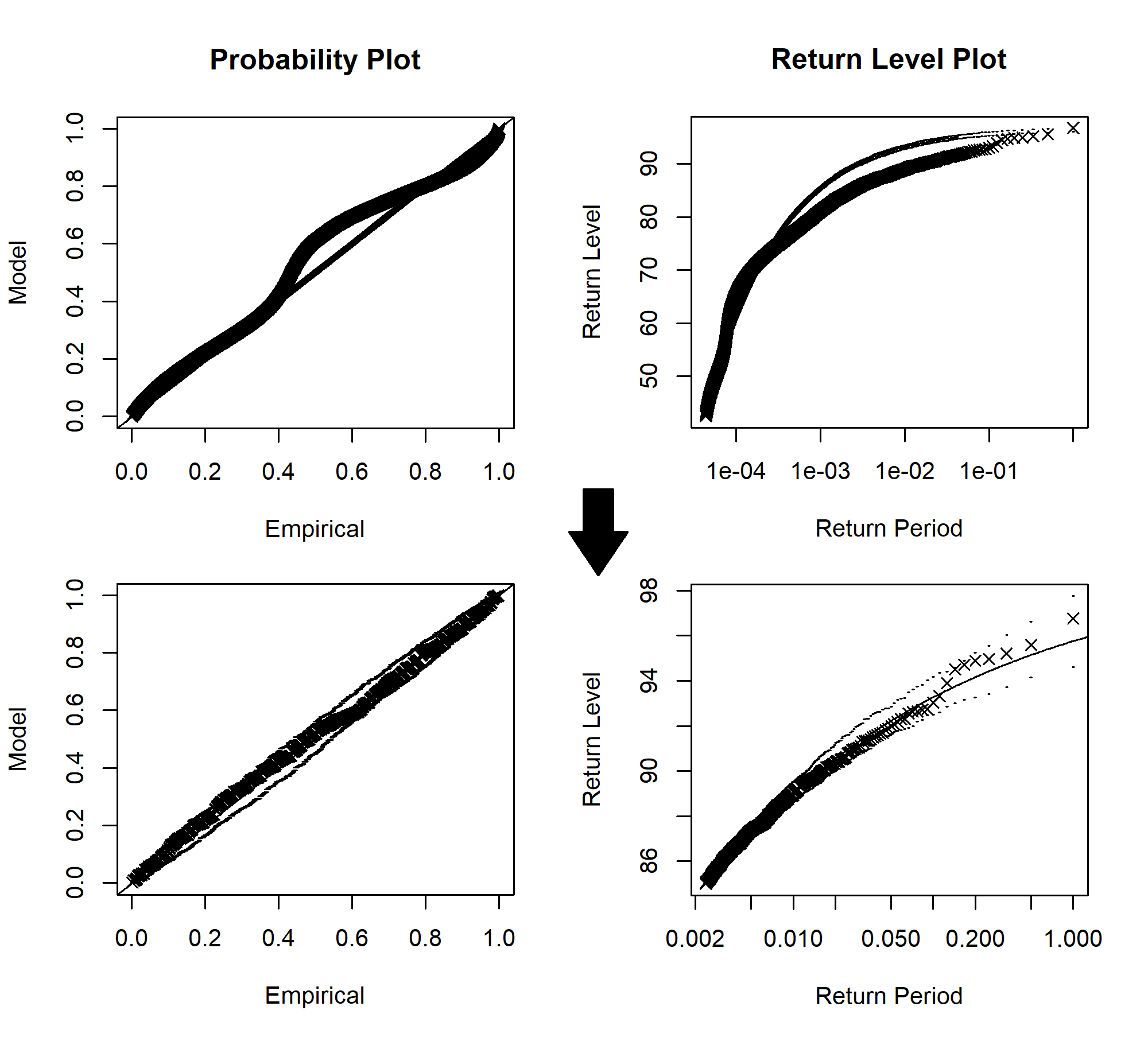}%
\label{fig:GOF_threshold}}
\hfil
\subfloat[Training performance]{\includegraphics[width=3.3in,valign=t, trim = {0cm 0cm 1cm -0.9cm}]{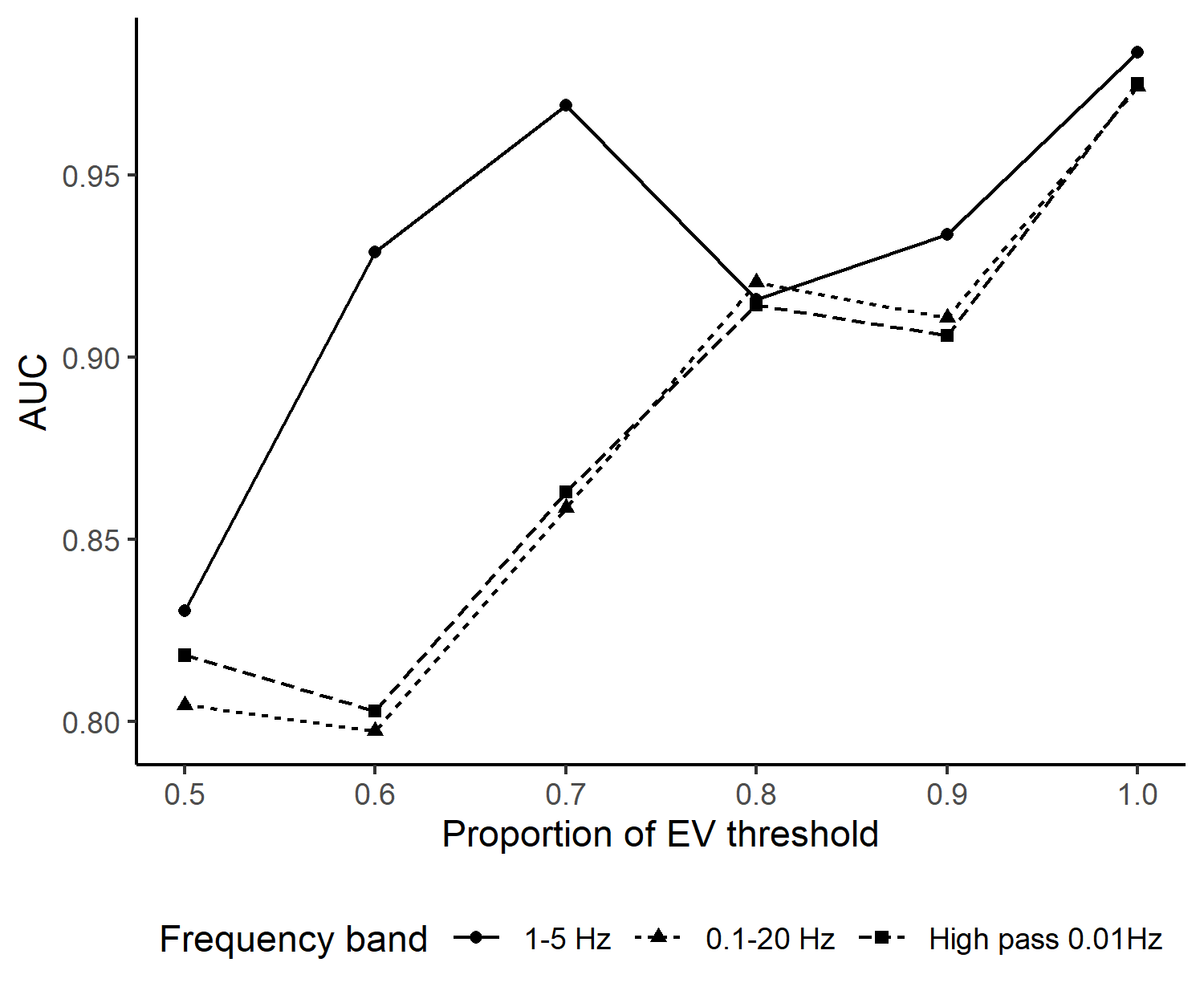}%
\label{fig:threshold_AUC}}

\caption{(a) Improvement in goodness-of-fit when the threshold used increases from 50\% (top) to 100\% (bottom) of the value chosen by extreme value theory; (b) Improvement in the training performance in terms of Area under the Curve (AUC) for the 1-5Hz, 0.1-20Hz and high pass 0.01Hz index exceedance models when the threshold increases.}
\label{fig_EVT_value}
\end{figure*}

\subsection{Results}

The black line in Figure \ref{fig:results_15_all} show the one-hour ahead probabilistic forecasts for threshold exceedances of the 1-5Hz envelope. The forecasts are highest for January 2nd, the day the eruption started. Focusing on January 2nd in Figure \ref{fig:results_15_zoom}, the exceedance probability jumps about 1 hour before the recorded eruption onset at 10:20. This means that fitted forecast model is able to give about 1 hour ahead warning before the eruption.
\\
Similar to \cite{Bee2019}, we check the goodness of fit of the logistic regression using a deviance chi-squared test. The p-value was $0$, indicating that the fitted model is significantly different from a null model. The usefulness of the covariates for explaining the temporal dependence in the occurrences of the threshold exceedances can also be seen through the reduction in autocorrelation of the Pearson residuals in Figure S1a of the Supplementary Information. In contrast, little temporal dependence was observed for the excess residuals in Figure S1b. Hence, there was no real benefit of using covariates to inform a dynamic GPD and a constant GPD would have sufficed. As we will see later, this is threshold-specific: when we use multiple events to train our model in Section \ref{sec:forecasteval} and select the lowest EVT-informed thresholds among the training events, there will be autocorrelation in the excess residuals and hence the benefits of modelling with a dynamic GPD. 
%here was difficulty in using Hosmer-Lemshow test for the null hypothesis of equality between the expected and observed frequency of exceedance. For the common choice of ten quantile groups, not all entries in the expected frequencies table were greater than 1 to ensure that the chi-squared approximation is valid. \\
%We check the goodness of fit of the GPD regression by comparing the predicted mean excess and the observed excesses, and comparing the quantiles of the standardised excesses to unit exponential quantiles. This is shown in Figure 1 of the Supplementary Information (SI). Poor fit suggests that a constant GPD fares better than a dynamic one (see more on this in the next Section).

\section{Value of extreme value theory} \label{sec:EVTvalue}

\begin{figure*}[!t]
\centering
\subfloat[GPD regression]{\includegraphics[width=2.5in,valign=t, trim = {0.2cm 0.2cm 0.2cm 0.2cm}]{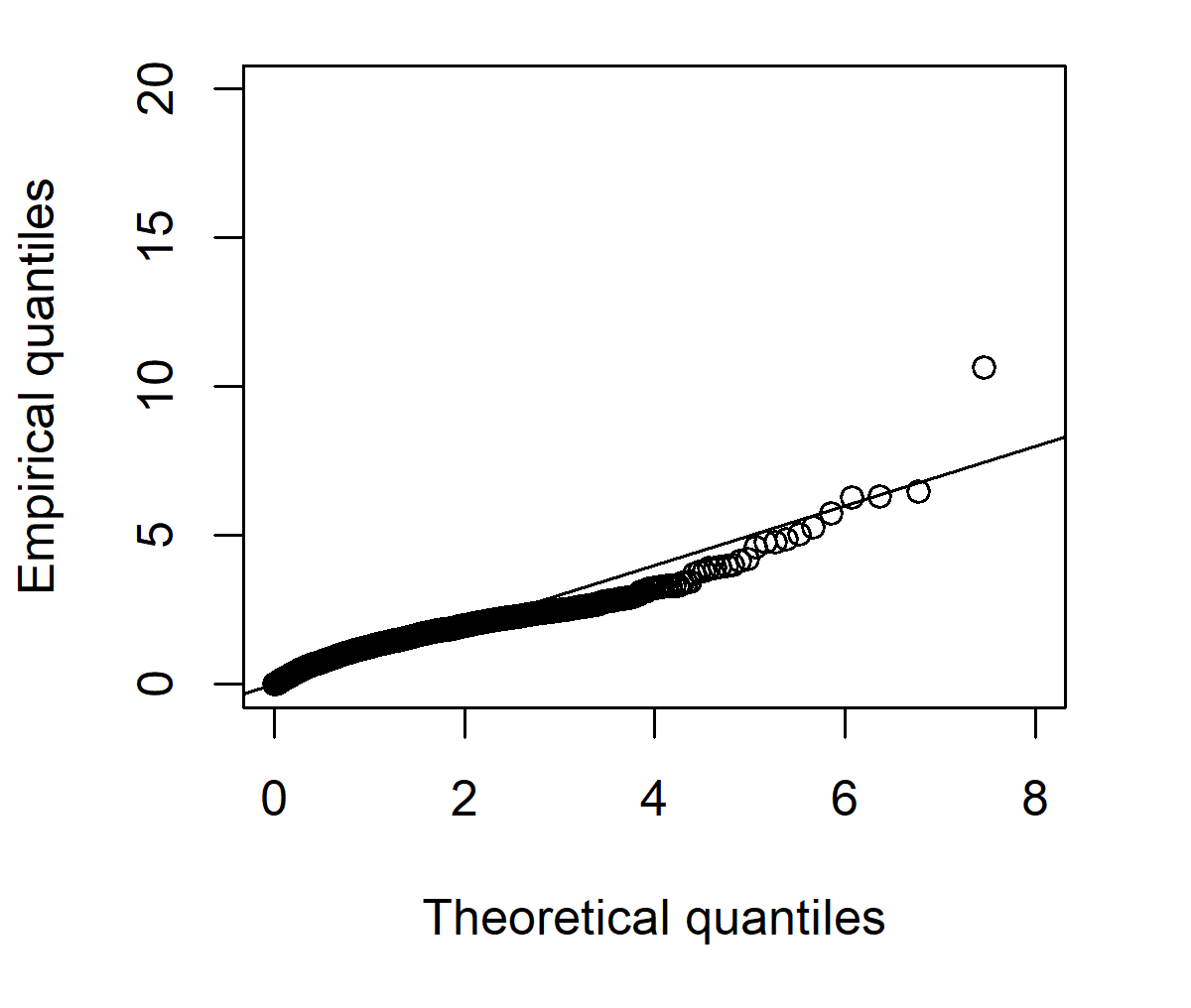}%
\label{fig:qq_GPDreg}}
\hfil
\subfloat[Constant GPD]{\includegraphics[width=2.5in,valign=t, trim = {0.2cm 0.2cm 0.2cm 0.2cm}]{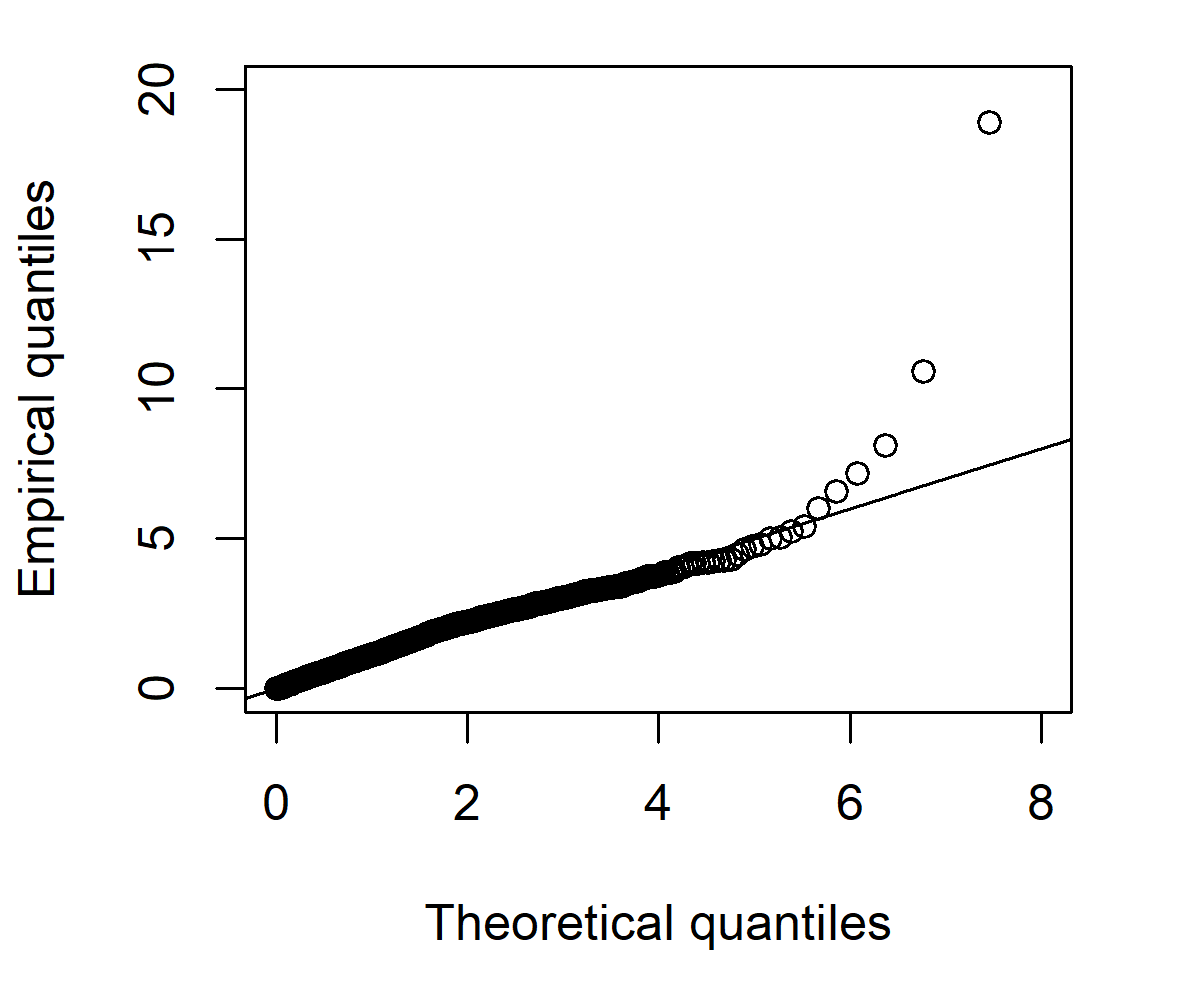}%
\label{fig:qq_cGPD}}

\caption{With multiple training events: Comparison in goodness-of-fit in terms of the empirical and theoretical quantiles of the standardised excesses when we use (a) the GPD regression for the excess distributions instead of (b) treating the excess distribution as static through a constant GPD. Since the extreme quantiles lie closer to the one-to-one diagonal line for the GPD regression, it fits the data better.}
\label{fig_qq_GPD}
\end{figure*}

EVT was used to select the threshold which defined the exceedances and excesses being forecasted by the dynamic extreme value model. Figure \ref{fig:GOF_threshold} shows the goodness-of-fit plots comparing the modelled and empirical probabilities and return levels when 50\% and 100\% of the threshold informed by EVT was used. The latter provided a better fit to the model assumptions. 
\\
The threshold choice is also important for determining what kind of phenomena is being modelled and what covariates are chosen to best explain it. Figure \ref{fig:threshold_AUC} shows that for the exceedance forecasting of the 1-5Hz, 0.1-20Hz and high pass 0.01Hz envelope indices, the forecast performance, as measured via Area Under the Curve (AUC), generally improves as we increase the threshold towards that informed by EVT. 
\\
Hence EVT has benefits for modelling in terms of both goodness of fit to the data and forecast performance.

\begin{figure*}[!t]
\centering
\subfloat[Training event 1]{\includegraphics[width=3.5in,valign=t]{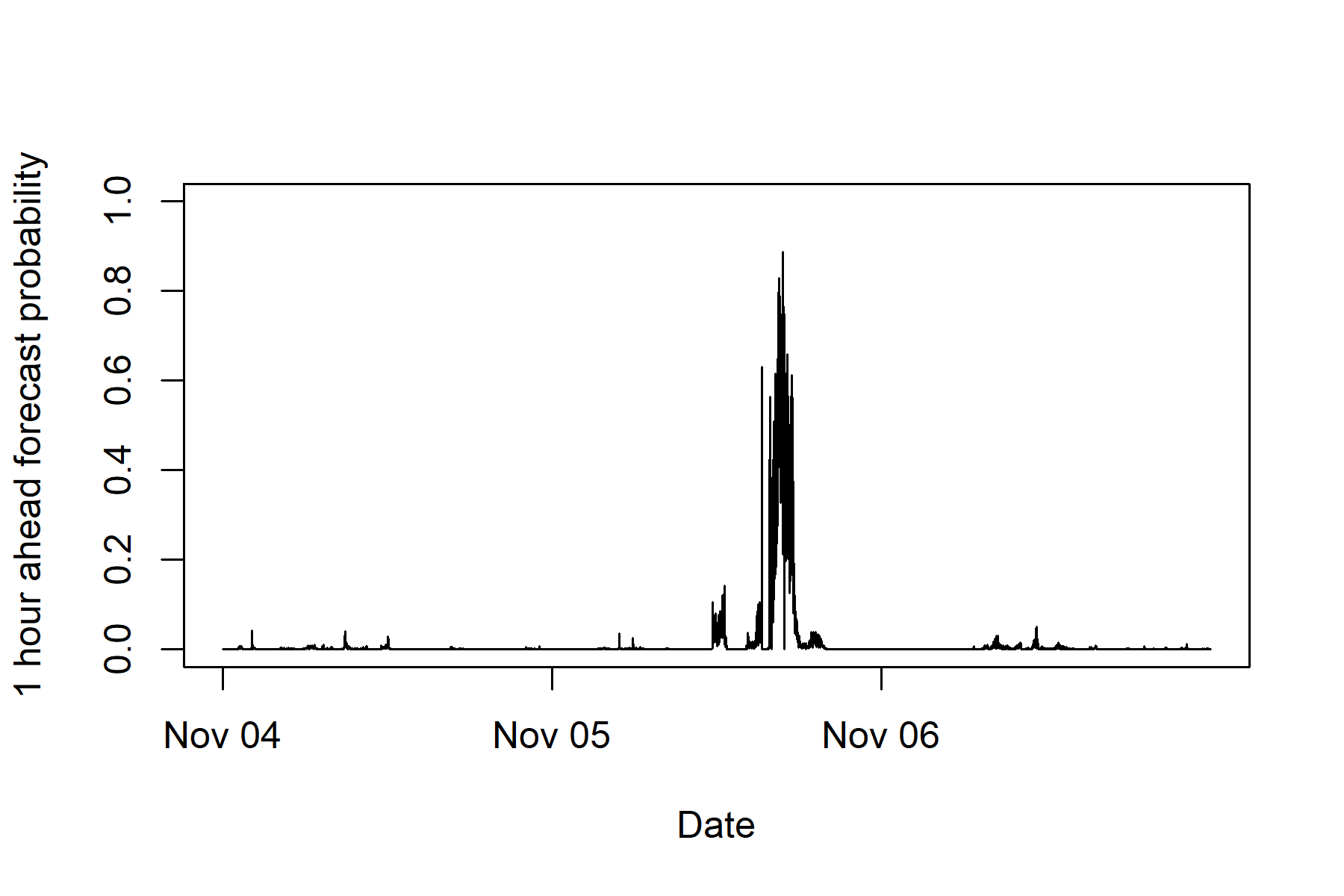}
\label{fig_tf_1}}
\subfloat[Zoomed in]{\includegraphics[width=3.5in,valign=t]{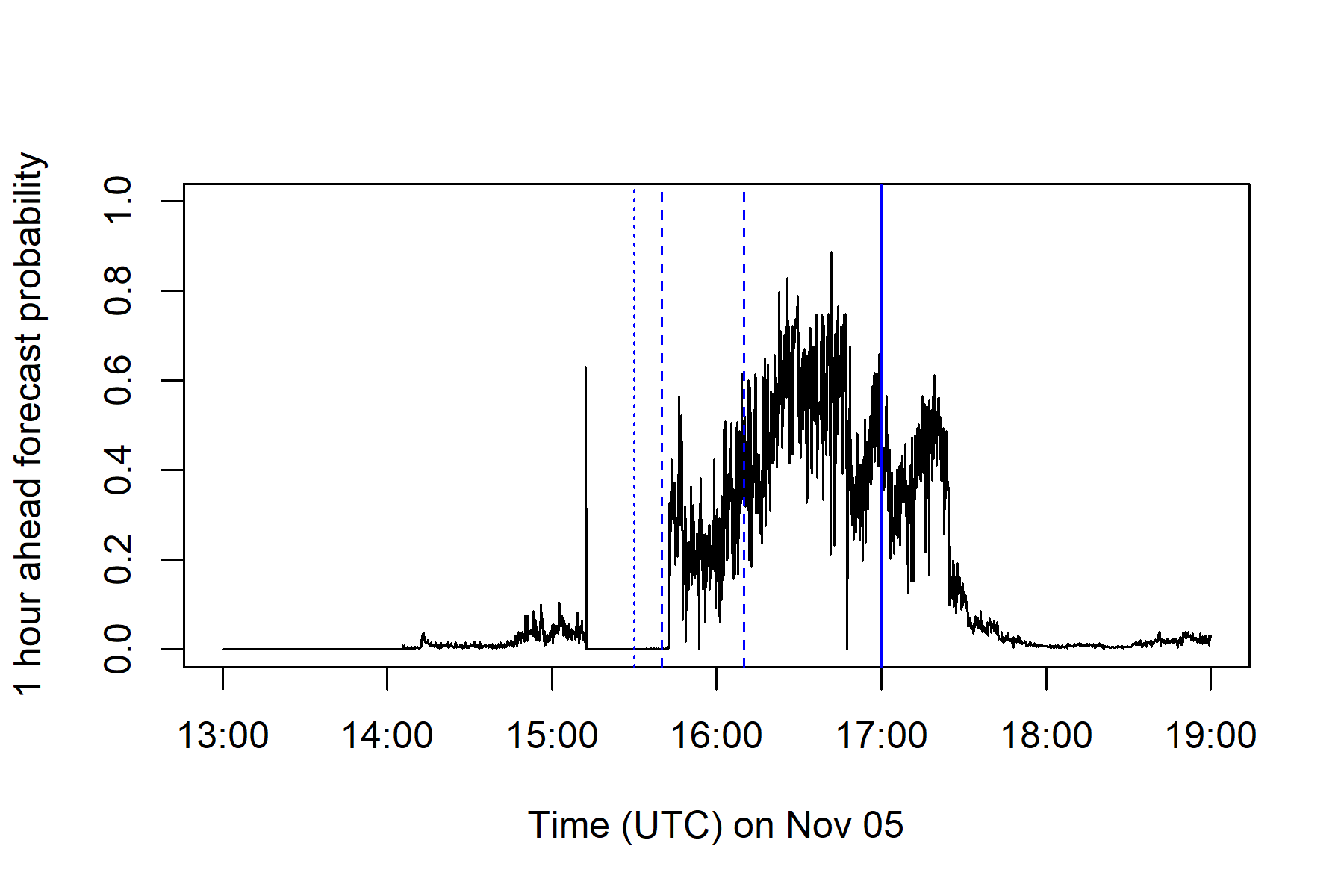}
\label{fig_tf_1_zoom}}
\\
\subfloat[Training event 2]{\includegraphics[width=3.5in,valign=t]{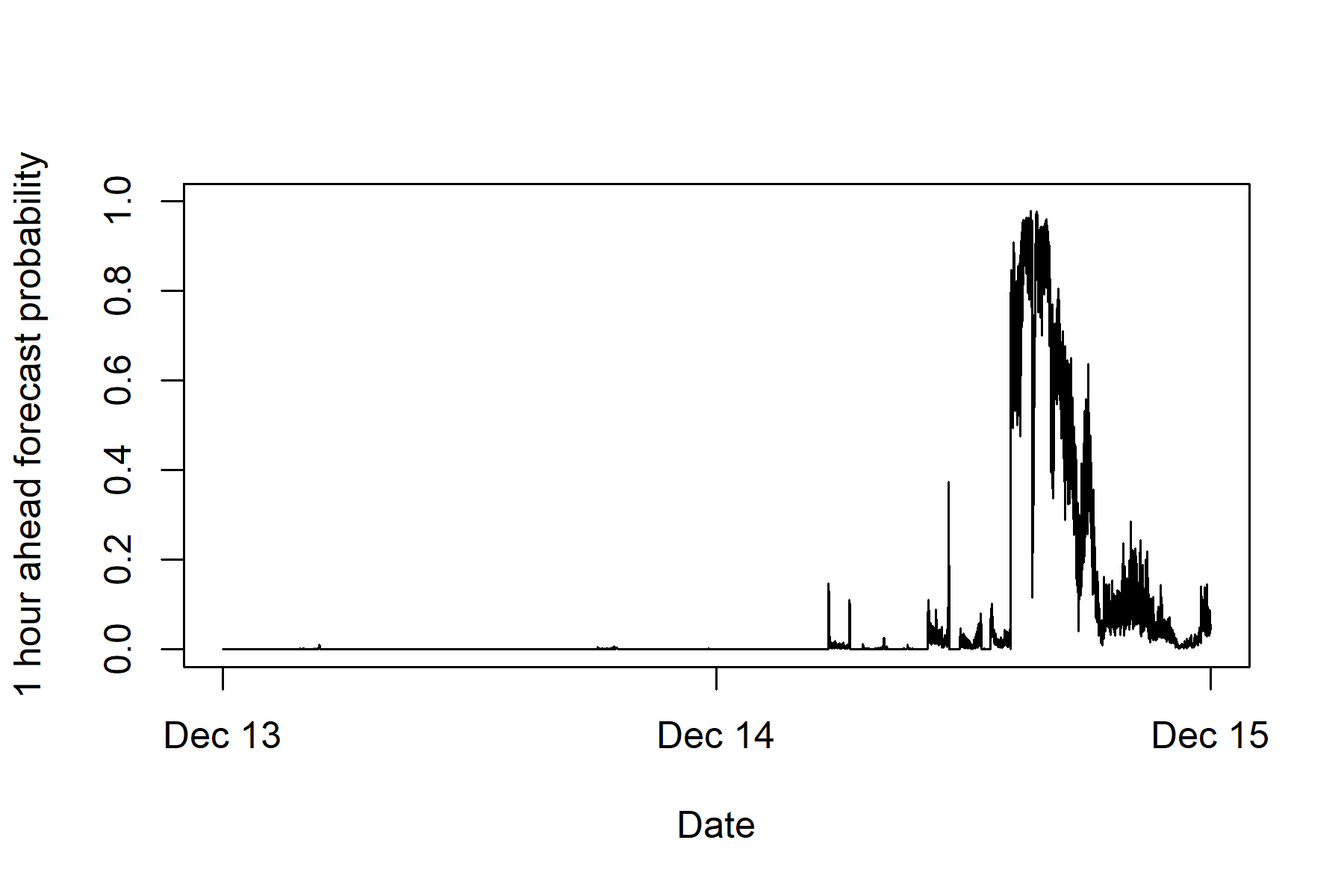}
\label{fig_tf_2}}
\subfloat[Zoomed in]{\includegraphics[width=3.5in,valign=t]{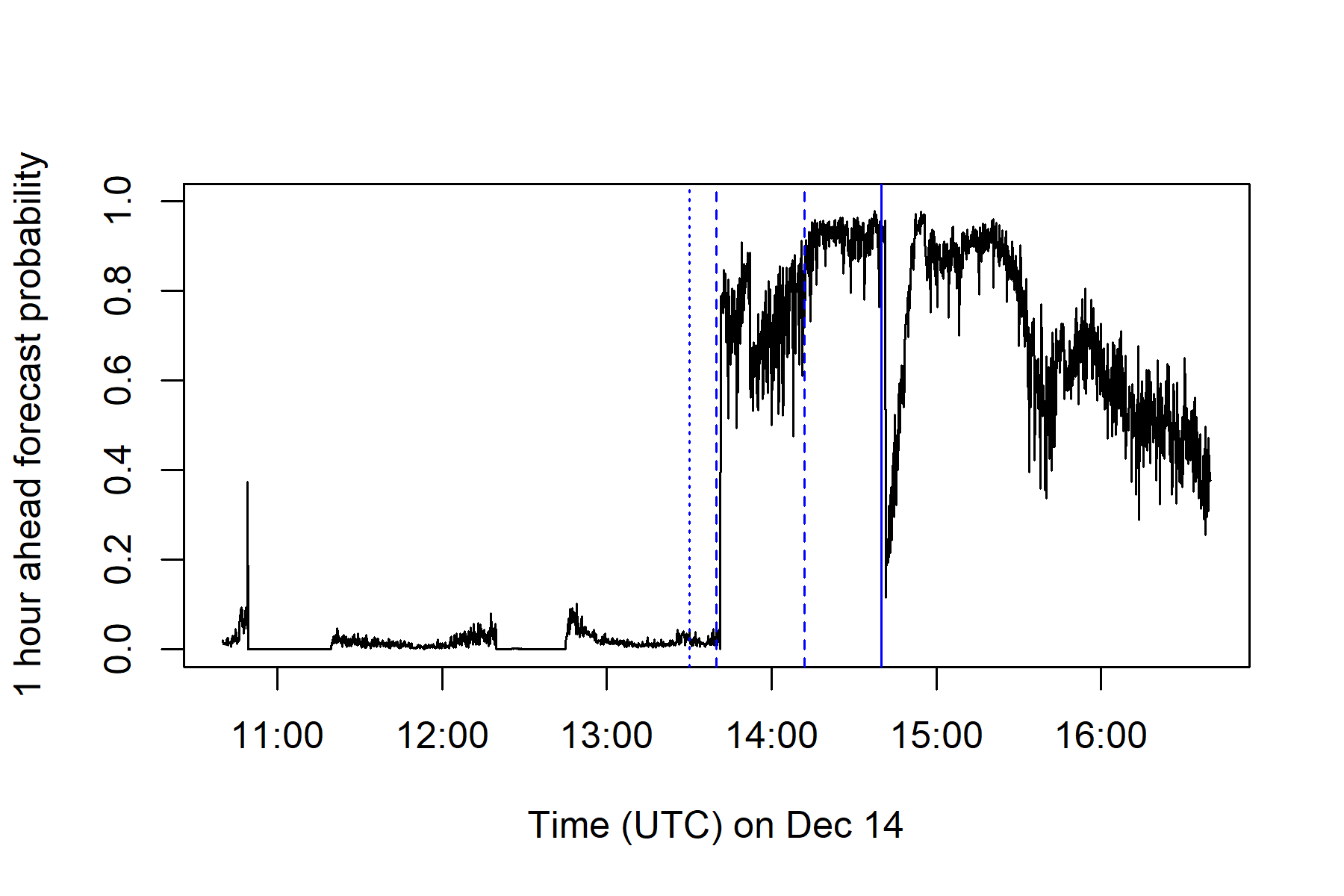}
\label{fig_tf_2_zoom}}
\\
\subfloat[Training event 3]{\includegraphics[width=3.5in,valign=t]{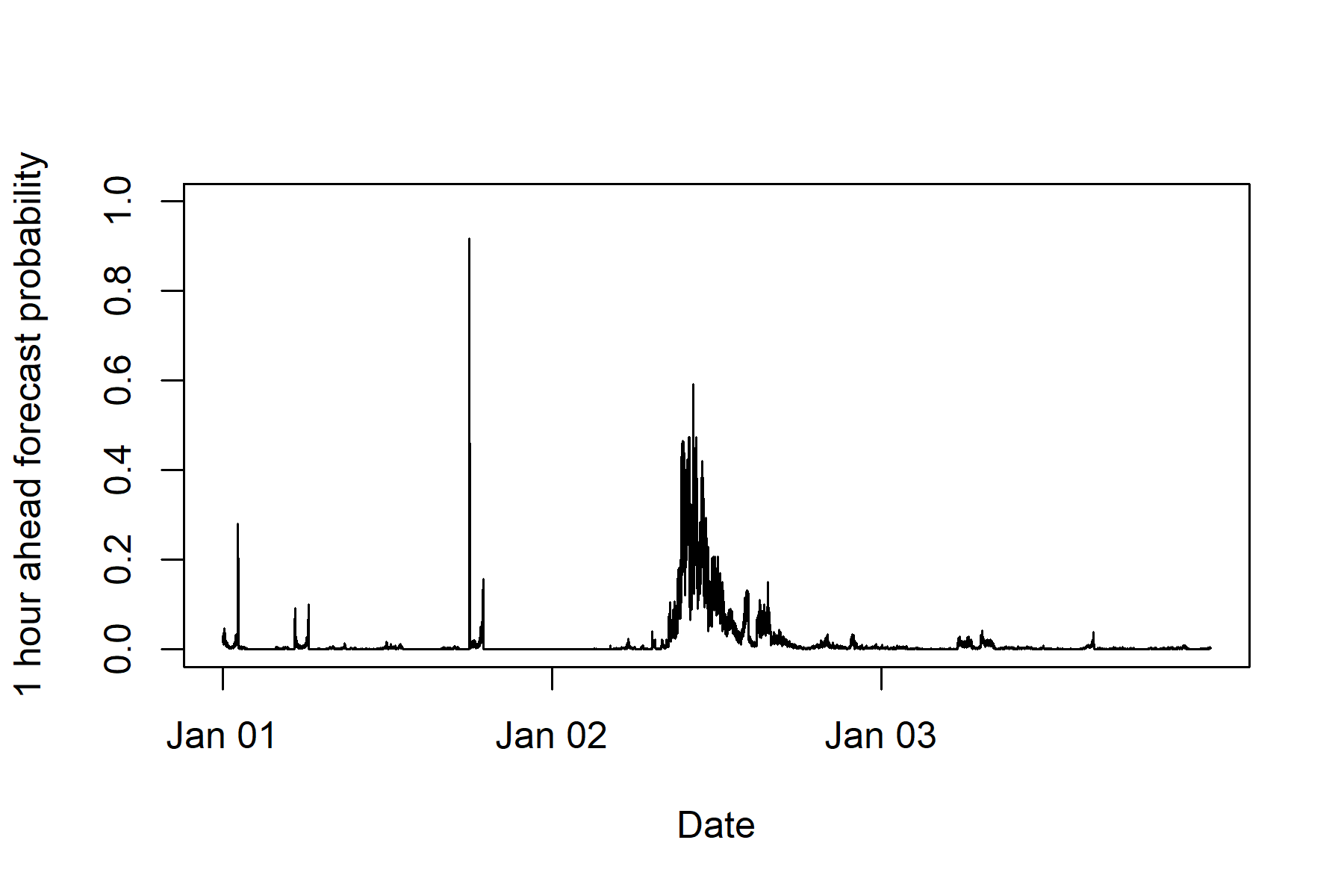}
\label{fig_tf_3}}
\subfloat[Zoomed in]{\includegraphics[width=3.5in,valign=t]{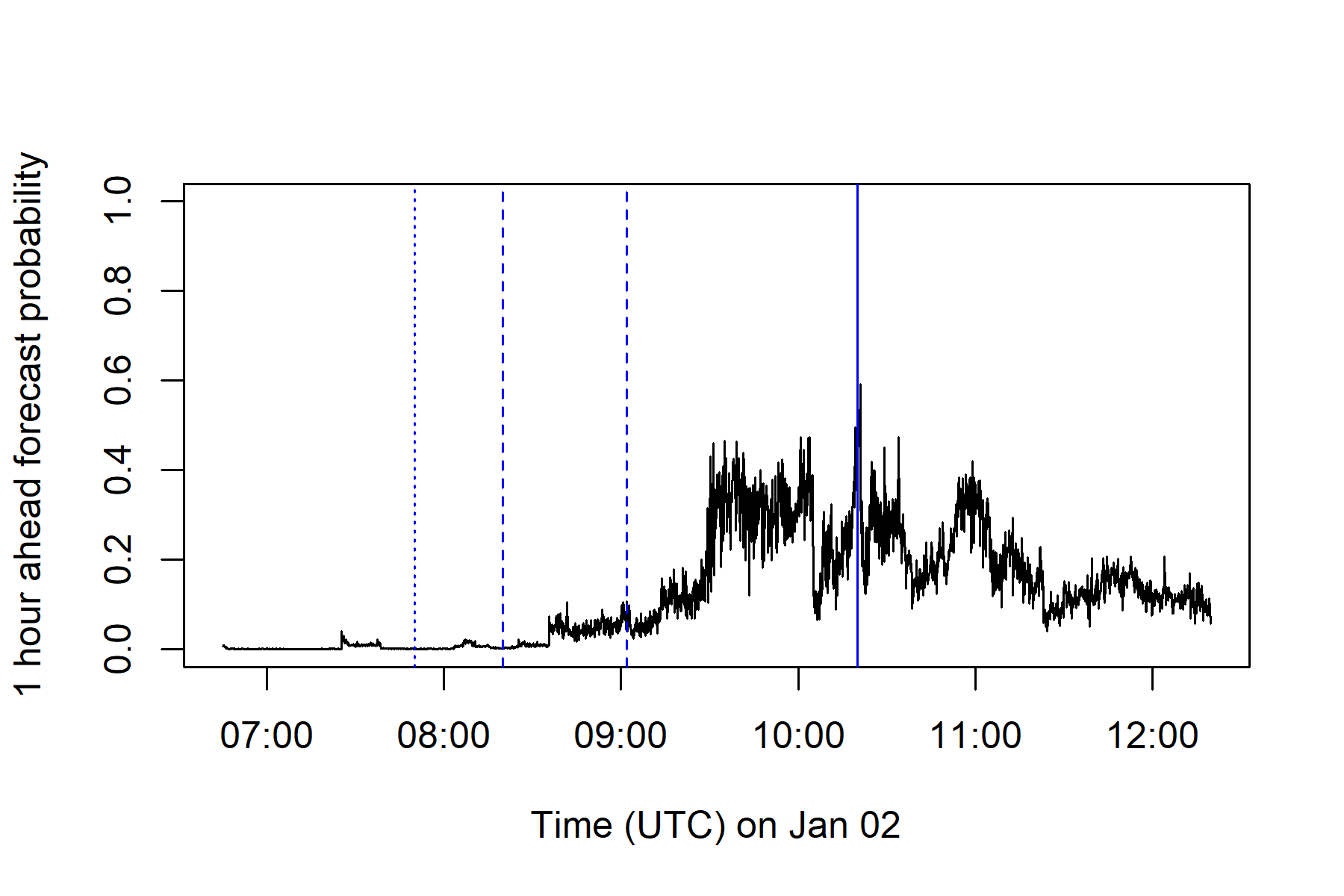}
\label{fig_tf_3_zoom}}

\caption{One-hour ahead threshold exceedance forecasts for the three training events using the lowest threshold estimated across events. The dotted, dashed and bold blue vertical lines denote the times of the seismic crises, start and end of the seismic swarms and the eruption onset respectively.}
\label{fig:training_forecasts}
\end{figure*}

\begin{figure*}[!t]
\centering
\subfloat[Test event]{\includegraphics[width=3.5in,valign=t]{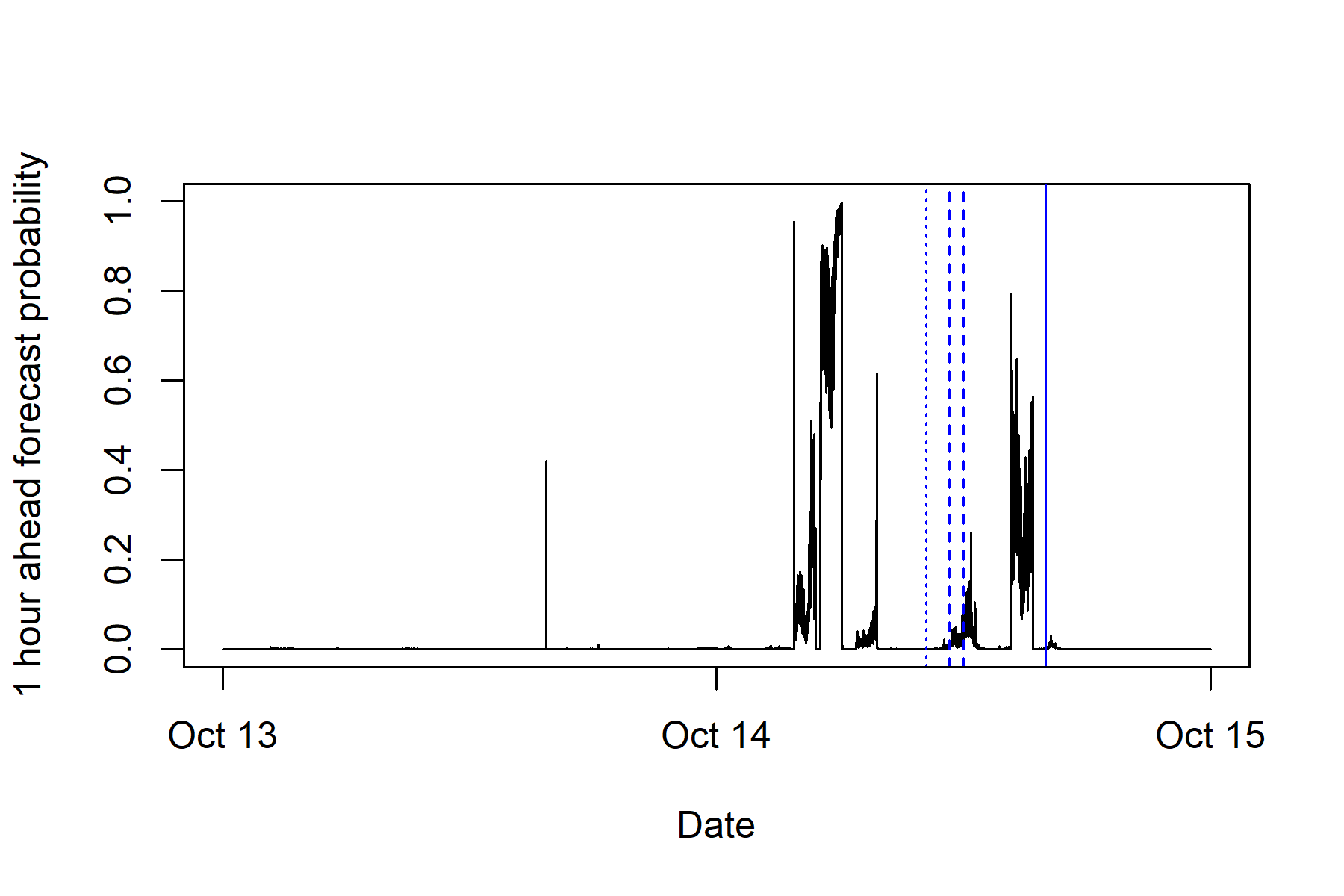}
\label{fig_test}}
%\subfloat[Zoomed in]{\includegraphics[width=3.5in,valign=t]{figures/te_prob_15_test_zoom_lbound.png}
%\label{fig_test_zoom}}
\subfloat[Test non-event 1]{\includegraphics[width=3.5in,valign=t]{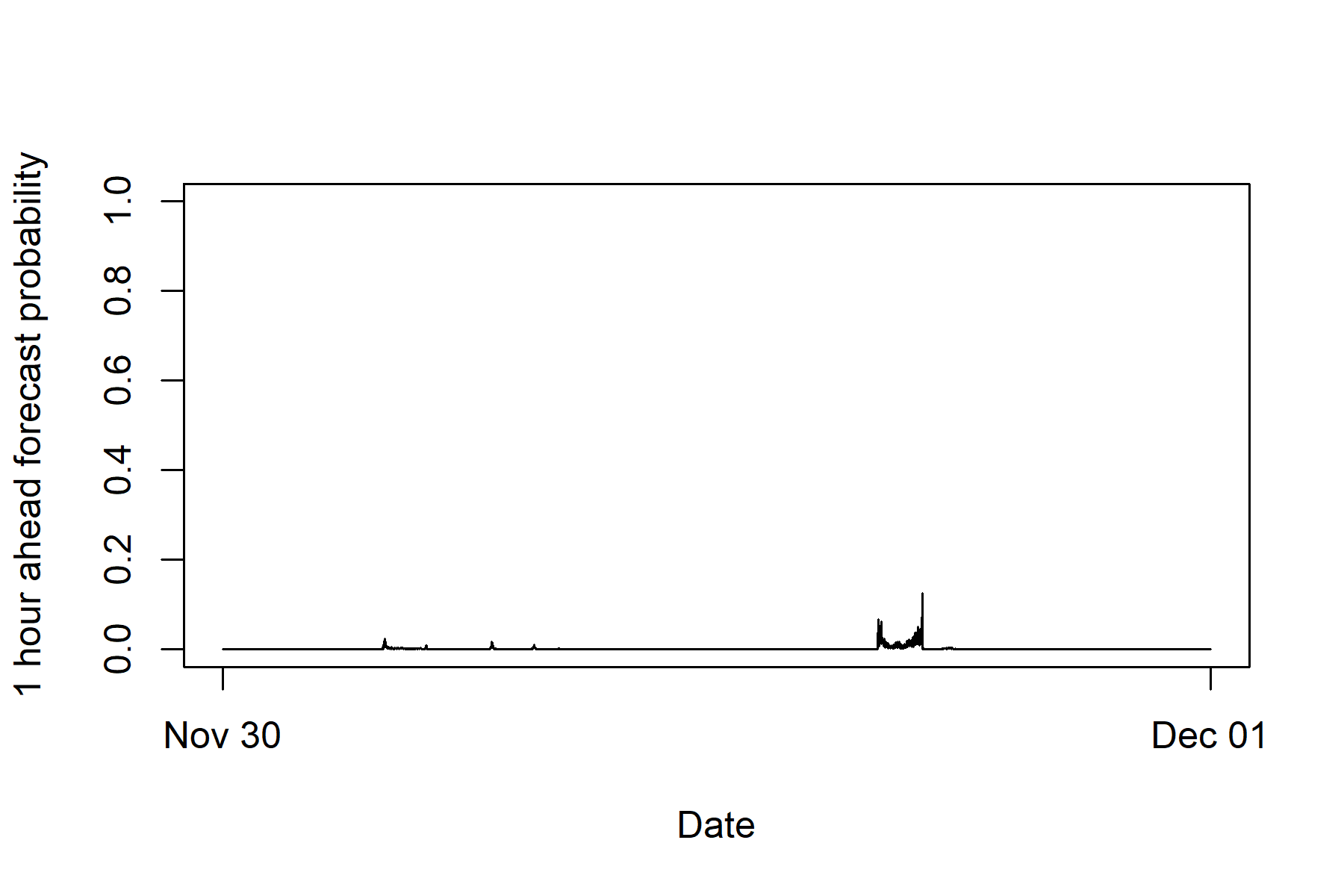}
\label{fig_nonevent_1}}
\\
\subfloat[Test non-event 2]{\includegraphics[width=3.5in,valign=t]{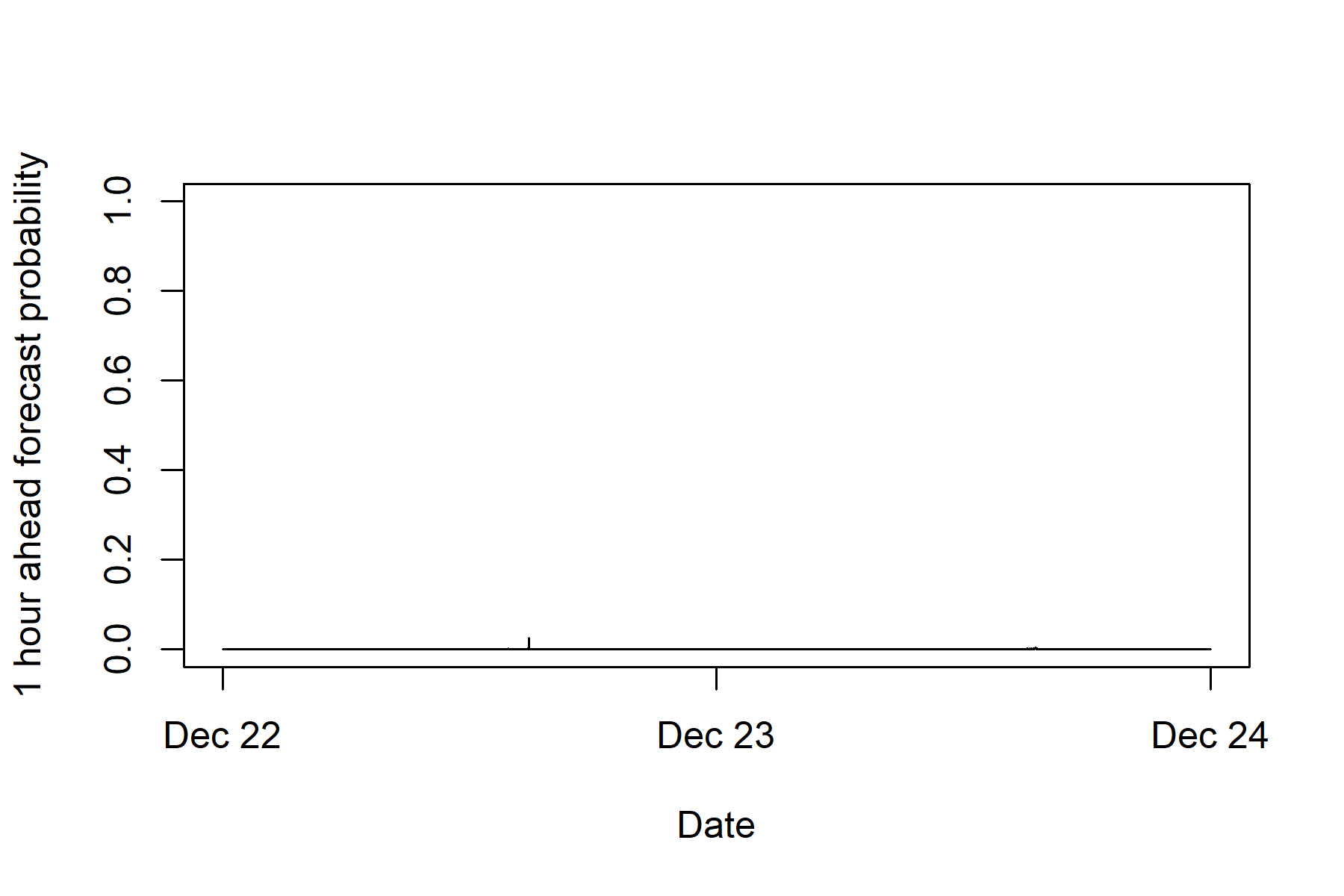}
\label{fig_nonevent_2}}
\subfloat[Test non-event 3]{\includegraphics[width=3.5in,valign=t]{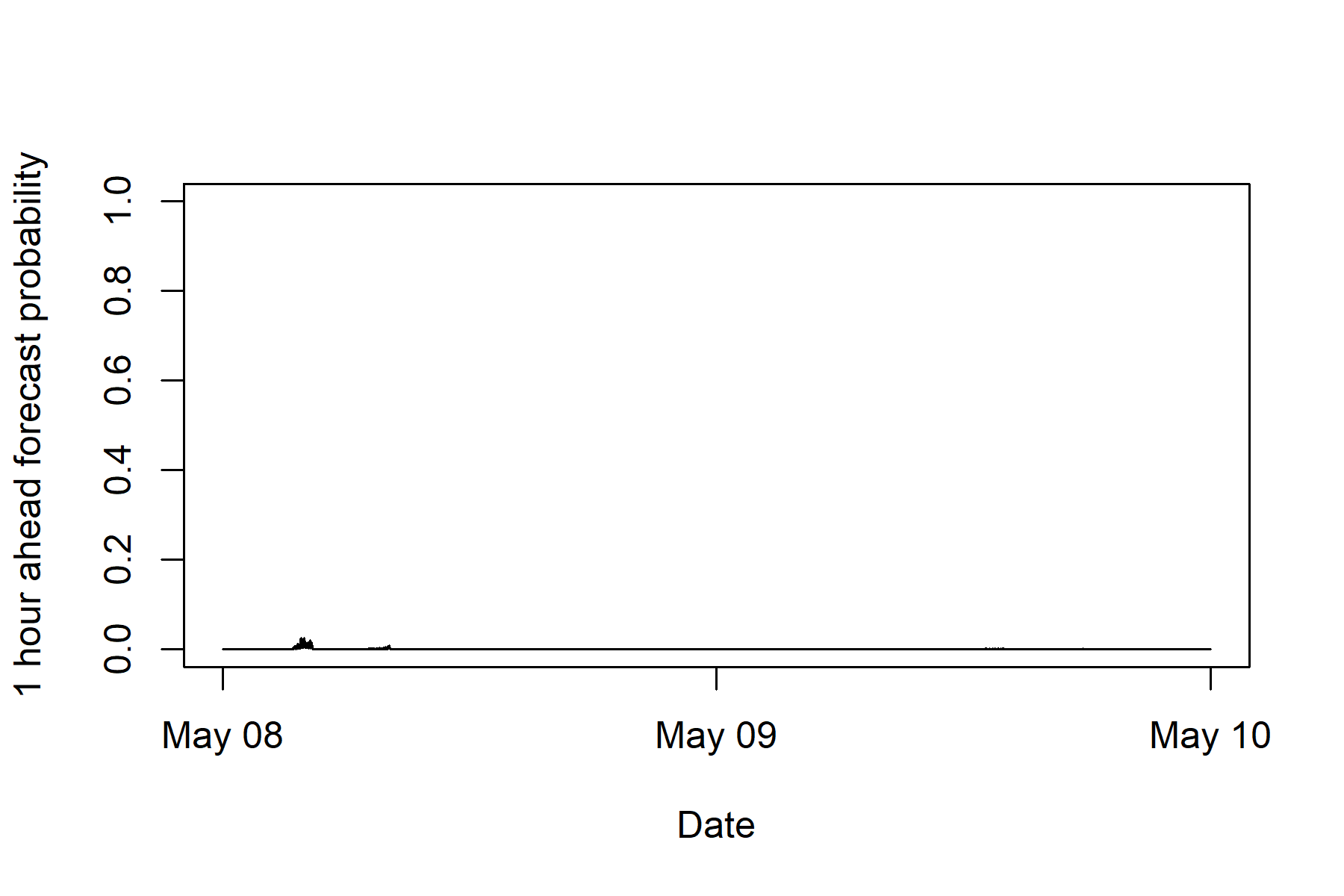}
\label{fig_nonevent_3}}
\caption{One-hour ahead threshold exceedance forecasts using the lowest threshold estimated across events for: (a) the test event, (b)-(d) the test non-events. The dotted, dashed and bold blue vertical lines in Plot (a) denote the times of the seismic crisis, seismic swarm and the eruption onset.}
\label{fig:test_forecast}
\end{figure*}

\section{Evaluating forecast performance} \label{sec:forecasteval}

\subsection{Using multiple training events}

To assess the forecast capability of the dynamic extreme value model more formally, we will fit the model to all three training events and test on the remaining test event and the three non-events. 
\\
There are a few ways we could determine a suitable threshold based on data from multiple events. An initial approach might be to simply combine the data across events and use all exceedances to inform the threshold. However, in our analysis, this led to a relatively high threshold estimate of 95 with Training event 2 dominating the model fit because there were comparatively more exceedances from Event 2 than Events 1 and 3 (see Section IV of the Supplementary Information). 
\\
An alternative approach, which we will use, would be to estimate thresholds for the training events separately and use the lowest estimate across events. This will ensure that we have sufficient exceedances to represent each event. For the 1-5Hz envelope index, the lowest threshold among the three training events was 85 and the estimated GPD shape parameter is $\hat{\xi} = -0.125$ with 95\% confidence interval $(-0.146, -0.104)$. Tables I and II of the Supplementary Information show the chosen covariates with their transformations and parameter estimates for the fitted logistic and GPD regressions respectively. 
\\
As we will see in the next Section, this threshold choice leads to reasonable one-hour ahead forecast probabilities for all three training events, the test event and the three test non-events. By lowering the threshold to identify extremes across all three training events, we also see more benefit of modelling the threshold excesses dynamically because there is autocorrelation in the excess residuals, particularly for Training event 2 (see Section II of the Supplementary Information). In contrast to our initial illustration with just Training Event 3 in Section \ref{sec:casestudy}, we see that modelling the excess distribution dynamically leads to better estimation of extreme quantiles, though there is still room for improvement. This is illustrated in Figure \ref{fig_qq_GPD}. 

\subsection{Training and test performance}

After fixing the threshold for which to model exceedances, we fit our dynamic extreme value model to data from the three training events. Figure \ref{fig:training_forecasts} shows that apart from an outlier on the first day of Training event 3, the forecast probabilities remain low (e.g. below 0.3) for all three events until the time of their recorded seismic events. For Training event 1 (referring to Figure \ref{fig_tf_1_zoom}), sustained high forecast probabilities start during the seismic swarm (between the dashed vertical lines), slightly more than one hour before the recorded eruption at 17:00. For Event 2, one hour ahead eruption warnings can also be made as the forecast probabilities begin to take high values during the seismic swarm, about an hour before the recorded eruption at 14:40 (see Figure \ref{fig_tf_2_zoom}). For Event 3, the forecast probabilities gradually increase from the time of the seismic swarm around 08:30 before jumping up to a higher plateau about an hour before the recorded eruption onset at 10:20 as shown in Figure \ref{fig_tf_3_zoom}. 
\\
The features of the forecast probabilities, namely the sharp jumps from near-zero for Training events 1 and 2, the gradual increase for Training event 3, the presence of outliers and the tendency to increase one hour before the recorded eruption onsets, stem from the chosen covariates of the logistic regression.  As can be inferred from their high coefficient estimates in Table I of the Supplementary Information, the logistic regression for threshold exceedance has three covariates which contribute to forecast probabilities more than the others: 0.1-20Hz cepstral kurtosis, 0.1-20Hz cepstral skewness and high pass 0.01Hz energy. 
\\
Figures S12-S15 in the Supplementary Information show the time series of these covariates for the training and test events. Unlike the 0.1-20Hz cepstral kurtosis and skewness, the high pass 0.01Hz energy has more block-like features in its time series. This influences the sharp jumps from near-zero for Training events 1 and 2 during their seismic swarms. In constrast, the change in high pass 0.01Hz energy during the period of the seismic events on January 2nd of Training event 3 was more smooth, resulting in a smoother increase in forecast probabilities. The block-like features result from one extreme value in the high pass 0.01Hz signal which causes high energy values for the length of the moving covariate window (1 hour). Since energy was defined as the sum of the squared signal, this covariate is also very sensitive to outliers. Future exploration can be done for making the covariates more robust to outliers.
\\
Still focusing on high pass 0.01Hz energy, we notice from Figures S12-15 that the covariate values tend to increase during the seismic swarms. Since the seismic swarms for Training events 1, 2 and 3 occur slightly more than one hour prior to their recorded eruptions, monitoring the energy values seems to provide good one-hour ahead forecasts for the eruptions. We hypothesise that the length of the covariate window and the forecast horizons can be optimised depending on the expected seismic crisis and swarm durations at a volcano. If the seismic swarms precede the eruption by a longer time period, as in the test event, the one hour ahead forecast probabilities may not be so temporally accurate. This is what we observe in Figure \ref{fig_test} for the one-hour ahead forecast probabilities for the test event. Here, we have a seismic crisis of 5 hours 30 minutes instead of the 1-2.5 hours durations for the training events. Referring to Figure S15 of the Supplementary Information, we see that the sole forecast probability outlier on October 13 and the spikes in forecast probabilities in the earlier part of October 14 are in line with spikes in the high pass 0.01Hz energy covariate. However, while the largest energy value was recorded during the seismic swarm (see Figure S15f), the forecast probability was higher nearer to the actual eruption onset. This shows the effect of the other covariates, including the 0.1-20Hz cepstral kurtosis and skewness, which work together to moderate the forecast probabilities. By nature of the covariates involved, the forecast probabilities are sensitive to different aspects of seismicity. 
\\
In addition to the training and test events, we examine the potential for the dynamic extreme value model for eruption forecasting by looking at its performance for the three non-events. In line with expectations, Figures \ref{fig_nonevent_1}-\ref{fig_nonevent_3} shows that the corresponding threshold exceedance forecasts remain very low, compared to the magnitudes during the training and test events. Similar training and test results were obtained for the 5-15Hz frequency-filtered data. The corresponding plots are given in Section VII of the Supplementary Information. 

\section{Discussion and outlook} \label{sec:discussion}

In Section \ref{sec:casestudy}, we fitted the dynamic extreme value model to Training Set 3, the seismic time series for the January 2010 eruption at Piton de la Fournaise. Promising results were obtained with spikes in the probabilistic forecasts about an hour prior to the eruption onset. This means that in this case one hour ahead warnings can be made with the chosen set-up: using the 1-5Hz trace envelope as an eruption index with a one hour covariate window and one hour forecast horizon. Similar performance was also seen when we fitted the model to more data in Section \ref{sec:forecasteval}. 
\\
In addition to the good forecast results, we saw the value of using EVT to choose the threshold for which we model exceedances. An appropriate threshold is important because it determines the balance between the amount of exceedances used to inform covariate selection and the adherence of the threshold excesses to the asymptotic theory. In general, with a higher threshold, we have less exceedances to inform the model which could mean higher estimation uncertainty but the excess distribution becomes closer to a GPD. This distribution is useful for computing extreme quantiles such as value at risk in the financial risk management.
\\
We showed in Section \ref{sec:casestudy} that using EVT to choose the threshold can improve forecast performance. When we increased the threshold towards its EVT-informed value, the AUC, a measure for how well the model can distinguish between exceedances and non-exceedances, generally increased for the 1-5Hz, 0.1-20Hz and high pass 0.01Hz envelope indices. Intuitively, the threshold determines what it means to be extreme and hence would affect the covariates selected and thus, the forecast performance. 
\\
In practice, one would aim to train the forecast model with as much relevant data as possible. In Section \ref{sec:forecasteval}, we demonstrated the considerations to make when incorporating data from different events. For example, we used the p-values of GPD goodness-of-fit tests to identify thresholds for which the excesses can be seen to follow a GPD. This procedure assumes that the same, constant GPD applies to all the training events. However, as we have observed, different events can suggest different thresholds. Specifically, for Training Event 2, we inferred a higher threshold for the energy-related envelope index. This difference in envelope index values between events could be because some eruptions occur closer to the measurement station or because the eruption itself has higher flux.
\\
Our proposed strategy to deal with the different threshold choices is to take the lowest identified threshold. The GPD regression component of the dynamic extreme value model would then model the non-stationarity of the GPD with appropriate covariates. In fact, modelling the excess distribution dynamically was seen to be more useful when we trained the model with multiple events and the lower threshold as compared to previously with just a single training event in Section \ref{sec:casestudy}. If instead, we chose the higher threshold, as suggested when we combined all the training data, we would forecast only large events because Training Event 2 would dominate the model fit, resulting in our inability to forecast Training Events 1 and 3 well.
\\
The proposed modelling framework can be extended in various ways. In our analysis, we used one hour covariate and forecast windows. More work can be done to optimise these durations which are likely to be related to the seismic crisis and swarm durations. While the training events had seismic crises which lasted about 1-2 hours, the test event had a much longer seismic crisis duration of 5h 30min. This could explain why the training forecasts were more temporally precise than the test forecasts.
\\
In our analysis, we focused on seismic signals in the 1-5Hz frequency range. As noted in the literature \cite{bormann2013} and observed from the AUC comparison in Figure  \ref{fig:threshold_AUC}, some frequency bands can be more useful for eruption forecasting than others. By combining information across useful frequency ranges through a joint, multivariate model, we can provide a more comprehensive eruption forecast.
\\
Similarly, we can incorporate different monitoring signals such as gas emissions and ground deformation in our model. So far, we have only used indices and covariates based on seismic signals due to the prevalence of seismometers as volcano monitoring tools. Future extensions can include different monitoring signals and account for their interdependence and shared covariates via joint models.
\\
Data from one seismic station (UV05) was used in our analysis. The promising results indicate that such methods can be useful for volcanoes where there is only one measuring station. Nevertheless, when we have multiple stations, we can extend the framework to model signals from the monitoring network as a whole and make fuller use of available data. Since distance from the eruption location is related to higher detected energies, varying forecast probabilities from different stations might help inform not just the timing, but also the location of the eruption.
\\
Given that the dynamic extreme value model has worked well for financial forecasting \cite{Bee2019} and we have shown that it can be adapted for volcanic eruption forecasting, we postulate that it has high potential to be further adapted for wider applications. In particular, high sampling-rate data were used in both the financial and volcanic contexts. For the former, it was used to compute realised variations over relatively short horizons while in the latter, it helped to separate different frequency bands of interest. High sampling-rate data relevant to other natural hazards and crises are also being made available; however, what is deemed as high sampling-rate is highly context-specific. For example, for sea levels, data were previously only publicly available at the monthly or annual scales. Hence, 1-15 minute resolutions are deemed as high sampling rate. These are increasingly sought after to study extreme sea levels and coastal flooding \cite{woodworth2016, ozsoy2016, zemunik2021}. Using such data, the dynamic extreme value model can be adapted to forecast extreme sea levels and their impact on coastal communities. While high sampling rates are good to have, the framework itself does not depend on this and in the absence of such data, we can still forecast, albeit on a coarser scale.
\\
To adapt the dynamic extreme value model to wider settings, there are several general considerations, namely: what is a suitable index to compute threshold exceedances of, what are reasonable forecast horizons and covariate windows, and what covariates can be used to inform future behaviour? One might also consider using algorithms for selecting the threshold automatically (see for example, \cite{Barder2018}).
\\
A general practical consideration which is shared across all contexts, be it finance, volcanoes or other hazards and crises is the translation of the forecast probabilities into decisive action. What forecast probability warrants a warning or more drastic measures such as evacuation? The optimal strategy may not be straightforward but may involve many competing priorities and constraints, and should be determined on a case-by-case basis with multiple stakeholders and not just by the modellers and scientists. 

\section*{Acknowledgment}

The data used for the analysis was collected by the Institut de Physique du Globe de Paris, Observatoire Volcanologique du Piton de la Fournaise (IPGP/OVPF) and the Laboratoire de Gèophysique Interne et Tectonophysique (LGIT) within the framework of ANR\_08\_RISK\_011/UnderVolc project. The sensors are properties of the réseau sismologique mobile Français, Sismob (INSU-CNRS). This work is supported by the NTU-Imperial collaboration grant INCF-2020-010 and the National Research Foundation, Prime Minister’s Office, Singapore under the NRF-NRFF2018-06 award. AV would like to acknowledge funding  by the Alan Turing
Institute-Lloyd’s Register Foundation Programme on Data-Centric Engineering.

% Can use something like this to put references on a page
% by themselves when using endfloat and the captionsoff option.
\ifCLASSOPTIONcaptionsoff
  \newpage
\fi

% trigger a \newpage just before the given reference
% number - used to balance the columns on the last page
% adjust value as needed - may need to be readjusted if
% the document is modified later
%\IEEEtriggeratref{8}
% The "triggered" command can be changed if desired:
%\IEEEtriggercmd{\enlargethispage{-5in}}

% references section

% can use a bibliography generated by BibTeX as a .bbl file
% BibTeX documentation can be easily obtained at:
% http://mirror.ctan.org/biblio/bibtex/contrib/doc/
% The IEEEtran BibTeX style support page is at:
% http://www.michaelshell.org/tex/ieeetran/bibtex/
\bibliographystyle{IEEEtran}
% argument is your BibTeX string definitions and bibliography database(s)
\bibliography{references}

% Generated by IEEEtran.bst, version: 1.14 (2015/08/26)
\begin{thebibliography}{10}
\providecommand{\url}[1]{#1}
\csname url@samestyle\endcsname
\providecommand{\newblock}{\relax}
\providecommand{\bibinfo}[2]{#2}
\providecommand{\BIBentrySTDinterwordspacing}{\spaceskip=0pt\relax}
\providecommand{\BIBentryALTinterwordstretchfactor}{4}
\providecommand{\BIBentryALTinterwordspacing}{\spaceskip=\fontdimen2\font plus
\BIBentryALTinterwordstretchfactor\fontdimen3\font minus
  \fontdimen4\font\relax}
\providecommand{\BIBforeignlanguage}[2]{{%
\expandafter\ifx\csname l@#1\endcsname\relax
\typeout{** WARNING: IEEEtran.bst: No hyphenation pattern has been}%
\typeout{** loaded for the language `#1'. Using the pattern for}%
\typeout{** the default language instead.}%
\else
\language=\csname l@#1\endcsname
\fi
#2}}
\providecommand{\BIBdecl}{\relax}
\BIBdecl

\bibitem{ribatet2016}
M.~Ribatet, ``Crash course on univariate extreme value theory,'' June 2016,
  lecture notes from the June 2016 Extreme Value Modeling and Water Resources
  summer school held in Lyon, France.

\bibitem{embrechts2013}
P.~Embrechts, C.~Kl{\"u}ppelberg, and T.~Mikosch, \emph{Modelling extremal
  events: for insurance and finance}.\hskip 1em plus 0.5em minus 0.4em\relax
  Springer Science \& Business Media, 2013, vol.~33.

\bibitem{beirlant2004}
J.~Beirlant, Y.~Goegebeur, J.~Segers, and J.~L. Teugels, \emph{Statistics of
  extremes: theory and applications}.\hskip 1em plus 0.5em minus 0.4em\relax
  John Wiley \& Sons, 2004, vol. 558.

\bibitem{Danielsson1997}
J.~Danielsson and C.~G. De~Vries, ``Tail index and quantile estimation with
  very high frequency data,'' \emph{Journal of empirical Finance}, vol.~4, no.
  2-3, pp. 241--257, 1997.

\bibitem{Longin2000}
F.~M. Longin, ``From value at risk to stress testing: The extreme value
  approach,'' \emph{Journal of Banking \& Finance}, vol.~24, no.~7, pp.
  1097--1130, 2000.

\bibitem{diebold1998}
F.~X. Diebold, T.~Schuermann, and J.~D. Stroughair, ``Pitfalls and
  opportunities in the use of extreme value theory in risk management,'' in
  \emph{Decision technologies for computational finance}.\hskip 1em plus 0.5em
  minus 0.4em\relax Springer, 1998, pp. 3--12.

\bibitem{Bee2019}
M.~Bee, D.~J. Dupuis, and L.~Trapin, ``{Realized peaks over threshold: A
  time-varying extreme value approach with high-frequency-based measures},''
  \emph{Journal of Financial Econometrics}, vol.~17, no.~2, pp. 254--283, 2019.

\bibitem{Malfante2018}
M.~Malfante, M.~{Dalla Mura}, J.~P. Metaxian, J.~I. Mars, O.~Macedo, and
  A.~Inza, ``{Machine Learning for Volcano-Seismic Signals: Challenges and
  Perspectives},'' \emph{IEEE Signal Processing Magazine}, vol.~35, no.~2, pp.
  20--30, 2018.

\bibitem{carniel2020}
\BIBentryALTinterwordspacing
R.~Carniel and S.~R. Guzmán, ``Machine learning in volcanology: A review,'' in
  \emph{Updates in Volcanology}, K.~Németh, Ed.\hskip 1em plus 0.5em minus
  0.4em\relax Rijeka: IntechOpen, 2020, ch.~5. [Online]. Available:
  \url{https://doi.org/10.5772/intechopen.94217}
\BIBentrySTDinterwordspacing

\bibitem{whitehead2021}
M.~G. Whitehead and M.~S. Bebbington, ``Method selection in short-term eruption
  forecasting,'' \emph{Journal of Volcanology and Geothermal Research}, p.
  107386, 2021.

\bibitem{Withers1998}
M.~Withers, R.~Aster, C.~Young, J.~Beiriger, M.~Harris, S.~Moore, and
  J.~Trujillo, ``A comparison of select trigger algorithms for automated global
  seismic phase and event detection,'' \emph{Bulletin of the Seismological
  Society of America}, vol.~88, no.~1, pp. 95--106, 1998.

\bibitem{Trnkoczy1999}
A.~Trnkoczy, ``Understanding and parameter setting of sta/lta trigger
  algorithm,''
  \url{https://gfzpublic.gfz-potsdam.de/rest/items/item_4097/component/file_4098/content},
  1999, [Online; accessed 27-April-2021].

\bibitem{Al-Mashhor2019}
A.~A. Al-Mashhor, A.~A. Al-Shuhail, S.~M. Hanafy, and W.~A. Mousa, ``First
  arrival picking of seismic data based on trace envelope,'' \emph{IEEE
  Access}, vol.~7, pp. 128\,806--128\,815, 2019.

\bibitem{sobradelo2015}
R.~Sobradelo and J.~Mart{\'\i}, ``Short-term volcanic hazard assessment through
  bayesian inference: retrospective application to the pinatubo 1991 volcanic
  crisis,'' \emph{Journal of Volcanology and Geothermal Research}, vol. 290,
  pp. 1--11, 2015.

\bibitem{bormann2013}
P.~Bormann, S.~Wendt, and K.~Klinge, ``Data analysis and seismogram
  interpretation,'' in \emph{New Manual of Seismological Observatory Practice 2
  (NMSOP-2)}.\hskip 1em plus 0.5em minus 0.4em\relax Deutsches
  GeoForschungsZentrum GFZ, 2013, pp. 1--151.

\bibitem{wild2021}
A.~J. Wild, M.~S. Bebbington, J.~M. Lindsay, and D.~H. Charlton, ``Modelling
  spatial population exposure and evacuation clearance time for the auckland
  volcanic field, new zealand,'' \emph{Journal of Volcanology and Geothermal
  Research}, vol. 416, p. 107282, 2021.

\bibitem{Sugihara1990}
G.~Sugihara and R.~M. May, ``Nonlinear forecasting as a way of distinguishing
  chaos from measurement error in time series,'' \emph{Nature}, vol. 344, no.
  6268, pp. 734--741, 1990.

\bibitem{Brenguier2008}
F.~Brenguier, N.~M. Shapiro, M.~Campillo, V.~Ferrazzini, Z.~Duputel,
  O.~Coutant, and A.~Nercessian, ``{Towards forecasting volcanic eruptions
  using seismic noise},'' \emph{Nature Geoscience}, vol.~1, no.~2, pp.
  126--130, 2008.

\bibitem{Roult2012}
\BIBentryALTinterwordspacing
G.~Roult, A.~Peltier, B.~Taisne, T.~Staudacher, V.~Ferrazzini, and A.~{Di
  Muro}, ``{A new comprehensive classification of the Piton de la Fournaise
  activity spanning the 1985-2010 period. Search and analysis of short-term
  precursors from a broad-band seismological station},'' \emph{Journal of
  Volcanology and Geothermal Research}, vol. 241-242, pp. 78--104, 2012.
  [Online]. Available: \url{http://dx.doi.org/10.1016/j.jvolgeores.2012.06.012}
\BIBentrySTDinterwordspacing

\bibitem{Taisne2011}
B.~Taisne, F.~Brenguier, N.~M. Shapiro, and V.~Ferrazzini, ``{Imaging the
  dynamics of magma propagation using radiated seismic intensity},''
  \emph{Geophysical Research Letters}, vol.~38, no.~4, pp. 2--6, 2011.

\bibitem{Journeau2020}
C.~Journeau, N.~M. Shapiro, L.~Seydoux, J.~Soubestre, V.~Ferrazzini, and
  A.~Peltier, ``{Detection, Classification, and Location of Seismovolcanic
  Signals with Multicomponent Seismic Data: Example from the Piton de La
  Fournaise Volcano (La R{\'{e}}union, France)},'' \emph{Journal of Geophysical
  Research: Solid Earth}, vol. 125, no.~8, pp. 1--19, 2020.

\bibitem{ren2020}
C.~X. Ren, A.~Peltier, V.~Ferrazzini, B.~Rouet-Leduc, P.~A. Johnson, and
  F.~Brenguier, ``Machine learning reveals the seismic signature of eruptive
  behavior at piton de la fournaise volcano,'' \emph{Geophysical research
  letters}, vol.~47, no.~3, p. e2019GL085523, 2020.

\bibitem{Barder2018}
B.~Barder, J.~Yan, and Z.~Xuebin, ``{Automated threshold selection for extreme
  value analysis via ordered goodness-of-fit tests with adjustment for false
  discovery rate},'' \emph{The Annals of Applied Statistics}, vol.~12, no.~1,
  pp. 310--329, 2018.

\bibitem{woodworth2016}
P.~L. Woodworth, J.~R. Hunter, M.~Marcos, P.~Caldwell, M.~Men{\'e}ndez, and
  I.~Haigh, ``Towards a global higher-frequency sea level dataset,''
  \emph{Geoscience Data Journal}, vol.~3, no.~2, pp. 50--59, 2016.

\bibitem{ozsoy2016}
O.~Ozsoy, I.~D. Haigh, M.~P. Wadey, R.~J. Nicholls, and N.~C. Wells,
  ``High-frequency sea level variations and implications for coastal flooding:
  A case study of the solent, uk,'' \emph{Continental Shelf Research}, vol.
  122, pp. 1--13, 2016.

\bibitem{zemunik2021}
P.~Zemunik, J.~{\v{S}}epi{\'c}, H.~Pellikka, L.~{\'C}atipovi{\'c}, and
  I.~Vilibi{\'c}, ``Minute sea-level analysis (misela): a high-frequency
  sea-level analysis global dataset,'' \emph{Earth System Science Data},
  vol.~13, no.~8, pp. 4121--4132, 2021.

\end{thebibliography}


% Generated by IEEEtran.bst, version: 1.14 (2015/08/26)
\begin{thebibliography}{10}
\providecommand{\url}[1]{#1}
\csname url@samestyle\endcsname
\providecommand{\newblock}{\relax}
\providecommand{\bibinfo}[2]{#2}
\providecommand{\BIBentrySTDinterwordspacing}{\spaceskip=0pt\relax}
\providecommand{\BIBentryALTinterwordstretchfactor}{4}
\providecommand{\BIBentryALTinterwordspacing}{\spaceskip=\fontdimen2\font plus
\BIBentryALTinterwordstretchfactor\fontdimen3\font minus
  \fontdimen4\font\relax}
\providecommand{\BIBforeignlanguage}[2]{{%
\expandafter\ifx\csname l@#1\endcsname\relax
\typeout{** WARNING: IEEEtran.bst: No hyphenation pattern has been}%
\typeout{** loaded for the language `#1'. Using the pattern for}%
\typeout{** the default language instead.}%
\else
\language=\csname l@#1\endcsname
\fi
#2}}
\providecommand{\BIBdecl}{\relax}
\BIBdecl

\bibitem{sigurdsson2015}
H.~Sigurdsson, B.~Houghton, S.~McNutt, H.~Rymer, and J.~Stix, \emph{The
  encyclopedia of volcanoes}.\hskip 1em plus 0.5em minus 0.4em\relax Elsevier,
  2015.

\bibitem{eggers1987}
A.~A. Eggers, ``Residual gravity changes and eruption magnitudes,''
  \emph{Journal of Volcanology and Geothermal Research}, vol.~33, no. 1-3, pp.
  201--216, 1987.

\bibitem{mcnutt1996}
S.~R. McNutt, ``Seismic monitoring and eruption forecasting of volcanoes: a
  review of the state-of-the-art and case histories,'' \emph{Monitoring and
  mitigation of volcano hazards}, pp. 99--146, 1996.

\bibitem{battaglia2003}
M.~Battaglia, P.~Segall, and C.~Roberts, ``{The mechanics of unrest at Long
  Valley caldera, California. 2. Constraining the nature of the source using
  geodetic and micro-gravity data},'' \emph{Journal of Volcanology and
  Geothermal Research}, vol. 127, no. 3-4, pp. 219--245, 2003.

\bibitem{burton2009}
M.~R. Burton, T.~Caltabiano, F.~Mur{\`e}, G.~Salerno, and D.~Randazzo, ``{SO2
  flux from Stromboli during the 2007 eruption: Results from the FLAME network
  and traverse measurements},'' \emph{Journal of Volcanology and Geothermal
  Research}, vol. 182, no. 3-4, pp. 214--220, 2009.

\bibitem{segall2013}
P.~Segall, ``Volcano deformation and eruption forecasting,'' \emph{Geological
  Society, London, Special Publications}, vol. 380, no.~1, pp. 85--106, 2013.

\bibitem{takahashi2015}
R.~Takahashi, T.~Shibata, Y.~Murayama, T.~Ogino, and N.~Okazaki, ``{Temporal
  changes in thermal waters related to volcanic activity of Tokachidake
  Volcano, Japan: implications for forecasting future eruptions},''
  \emph{Bulletin of Volcanology}, vol.~77, no.~1, pp. 1--12, 2015.

\bibitem{demoor2016}
J.~M. de~Moor, A.~Aiuppa, J.~Pacheco, G.~Avard, C.~Kern, M.~Liuzzo,
  M.~Martinez, G.~Giudice, and T.~P. Fischer, ``{Short-period volcanic gas
  precursors to phreatic eruptions: Insights from Po{\'a}s Volcano, Costa
  Rica},'' \emph{Earth and Planetary Science Letters}, vol. 442, pp. 218--227,
  2016.

\bibitem{poland2020}
M.~P. Poland, T.~Lopez, R.~Wright, and M.~J. Pavolonis, ``Forecasting,
  detecting, and tracking volcanic eruptions from space,'' \emph{Remote Sensing
  in Earth Systems Sciences}, vol.~3, no.~1, pp. 55--94, 2020.

\bibitem{marzocchi2012}
W.~Marzocchi, C.~Newhall, and G.~Woo, ``The scientific management of volcanic
  crises,'' \emph{Journal of Volcanology and Geothermal Research}, vol. 247,
  pp. 181--189, 2012.

\bibitem{whitehead2021}
M.~G. Whitehead and M.~S. Bebbington, ``Method selection in short-term eruption
  forecasting,'' \emph{Journal of Volcanology and Geothermal Research}, p.
  107386, 2021.

\bibitem{cooke1991}
R.~Cooke, \emph{Experts in uncertainty: opinion and subjective probability in
  science}.\hskip 1em plus 0.5em minus 0.4em\relax Oxford University Press on
  Demand, 1991.

\bibitem{aspinall2010}
W.~Aspinall, ``A route to more tractable expert advice,'' \emph{Nature}, vol.
  463, no. 7279, pp. 294--295, 2010.

\bibitem{colson2020}
A.~R. Colson and R.~M. Cooke, ``Expert elicitation: using the classical model
  to validate experts’ judgments,'' \emph{Review of Environmental Economics
  and Policy}, 2020.

\bibitem{newhall2002}
C.~Newhall and R.~Hoblitt, ``Constructing event trees for volcanic crises,''
  \emph{Bulletin of Volcanology}, vol.~64, no.~1, pp. 3--20, 2002.

\bibitem{marzocchi2004}
W.~Marzocchi, L.~Sandri, P.~Gasparini, C.~Newhall, and E.~Boschi,
  ``{Quantifying probabilities of volcanic events: the example of volcanic
  hazard at Mount Vesuvius},'' \emph{Journal of Geophysical Research: Solid
  Earth}, vol. 109, no. B11, 2004.

\bibitem{newhall2015}
C.~G. Newhall and J.~S. Pallister, ``Using multiple data sets to populate
  probabilistic volcanic event trees,'' in \emph{Volcanic Hazards, risks and
  disasters}.\hskip 1em plus 0.5em minus 0.4em\relax Elsevier, 2015, pp.
  203--232.

\bibitem{aspinall2003}
W.~P. Aspinall, G.~Woo, B.~Voight, and P.~J. Baxter, ``Evidence-based
  volcanology: application to eruption crises,'' \emph{Journal of Volcanology
  and Geothermal Research}, vol. 128, no. 1-3, pp. 273--285, 2003.

\bibitem{hill1977}
D.~P. Hill, ``A model for earthquake swarms,'' \emph{Journal of Geophysical
  Research}, vol.~82, no.~8, pp. 1347--1352, 1977.

\bibitem{chouet2003}
B.~Chouet, ``Volcano seismology,'' \emph{Pure and applied geophysics}, vol.
  160, no.~3, pp. 739--788, 2003.

\bibitem{fournier2007}
R.~O. Fournier, ``Hydrothermal systems and volcano geochemistry,'' in
  \emph{Volcano deformation}.\hskip 1em plus 0.5em minus 0.4em\relax Springer,
  2007, pp. 323--341.

\bibitem{salvage2016}
R.~Salvage and J.~Neuberg, ``{Using a cross correlation technique to refine the
  accuracy of the Failure Forecast Method: Application to Soufri{\`e}re Hills
  volcano, Montserrat},'' \emph{Journal of Volcanology and Geothermal
  Research}, vol. 324, pp. 118--133, 2016.

\bibitem{white2019}
R.~A. White and W.~A. McCausland, ``A process-based model of pre-eruption
  seismicity patterns and its use for eruption forecasting at dormant
  stratovolcanoes,'' \emph{Journal of Volcanology and Geothermal Research},
  vol. 382, pp. 267--297, 2019.

\bibitem{voight1988}
B.~Voight, ``A method for prediction of volcanic eruptions,'' \emph{Nature},
  vol. 332, no. 6160, pp. 125--130, 1988.

\bibitem{sparks2003}
R.~S. Sparks, ``{Forecasting volcanic eruptions},'' \emph{Earth and Planetary
  Science Letters}, vol. 210, no. 1-2, pp. 1--15, 2003.

\bibitem{bell2011}
A.~F. Bell, M.~Naylor, M.~J. Heap, and I.~G. Main, ``Forecasting volcanic
  eruptions and other material failure phenomena: an evaluation of the failure
  forecast method,'' \emph{Geophysical Research Letters}, vol.~38, no.~15,
  2011.

\bibitem{bell2013}
A.~F. Bell, M.~Naylor, and I.~G. Main, ``{The limits of predictability of
  volcanic eruptions from accelerating rates of earthquakes},''
  \emph{Geophysical Journal International}, vol. 194, no.~3, pp. 1541--1553,
  2013.

\bibitem{bevilacqua2019}
A.~Bevilacqua, E.~B. Pitman, A.~Patra, A.~Neri, M.~Bursik, and B.~Voight,
  ``Probabilistic enhancement of the failure forecast method using a stochastic
  differential equation and application to volcanic eruption forecasts,''
  \emph{Frontiers in Earth Science}, vol.~7, p. 135, 2019.

\bibitem{Malfante2018}
M.~Malfante, M.~{Dalla Mura}, J.~P. Metaxian, J.~I. Mars, O.~Macedo, and
  A.~Inza, ``{Machine Learning for Volcano-Seismic Signals: Challenges and
  Perspectives},'' \emph{IEEE Signal Processing Magazine}, vol.~35, no.~2, pp.
  20--30, 2018.

\bibitem{carniel2020}
\BIBentryALTinterwordspacing
R.~Carniel and S.~R. Guzmán, ``Machine learning in volcanology: A review,'' in
  \emph{Updates in Volcanology}, K.~Németh, Ed.\hskip 1em plus 0.5em minus
  0.4em\relax Rijeka: IntechOpen, 2020, ch.~5. [Online]. Available:
  \url{https://doi.org/10.5772/intechopen.94217}
\BIBentrySTDinterwordspacing

\end{thebibliography}
\end{document}